\documentclass[final,3p]{elsarticle}
\usepackage{times}
\usepackage{hyperref}	
\usepackage{anyfontsize}
\usepackage[utf8]{inputenc}
\usepackage{pdflscape} 
\usepackage{tabularx} 
\usepackage[british]{babel}
\usepackage[babel]{csquotes}
\usepackage{pbox}
\usepackage{float}
\usepackage{caption}
\usepackage{tcolorbox}
\usepackage{graphicx}
\usepackage{listings}
\usepackage{amssymb}
\usepackage{lineno}
\usepackage{lscape}
\usepackage{booktabs}
\usepackage{siunitx}
\newcolumntype{L}{@{}l@{}} 
\journal{Information and Software Technology}

\begin{document}

\begin{frontmatter}

\title{Comparing Open-Source and Commercial LLMs for Domain-Specific Analysis and Reporting: Software Engineering Challenges and Design Trade-offs}



\author[chalmers]{Theo Koraag}
\ead{}

\author[chalmers]{Niklas Wagner}
\ead{}

\author[miun]{Felix Dobslaw}
\ead{felix.dobslaw@miun.se}

\author[chalmers,getinge]{Lucas Gren\corref{cor1}}
\ead{lucas.gren@lucasgren.com}
\cortext[cor1]{Corresponding author}

\address[chalmers]{The Department of Computer Science and Engineering,\\ Chalmers University of Technology and The University of Gothenburg, Gothenburg, Sweden}
\address[miun]{Department of Communication, Quality Management and Information Systems,\\ Mid Sweden University, Östersund, Sweden}
\address[getinge]{Getinge AB, Gothenburg, Sweden}

\begin{abstract}
\textbf{Context:} The emergence of Large Language Models (LLMs) has enabled automation of complex natural language processing across a wide range of domains. However, research on the potential of different LLM solutions for specific domain-specific areas is scarce. One such domain is Finance. 

\textbf{Objectives:} This study explored the potential of open-source and commercial LLMs in financial report analysis and commentary generation, with a particular focus on the software engineering challenges encountered during the design and implementation of such solutions.

\textbf{Methods:} A Design Science Research methodology was used, where an exploratory case study was conducted. Two LLM-based systems were iteratively designed and evaluated: one employing local open-source models in a multi-agent workflow, and the other utilizing the commercial GPT-4o. Both solutions were evaluated through expert assessments of a real-world financial reporting use case.

\textbf{Results:} We observed that LLMs had great potential to automate tasks within financial reporting workflows, yet their integration presents challenges. Through iterative development and expert evaluation, several issues were identified, including prompt design, contextual dependency, and trade-offs between implementation options. Cloud-based models were found to offer greater fluency and ease of use, but raised concerns related to data privacy and reliance on external services. In contrast, local open-source models provide stronger data control and compliance but require substantially more engineering effort to ensure reliability and usability.

\textbf{Conclusion:} LLMs showed strong potential for automating financial reporting, but their integration requires careful attention to architecture, prompt design, and system reliability. Successful implementation depends on addressing domain-specific challenges through tailored validation mechanisms and engineering strategies that strike a balance between accuracy, control, and compliance.
\end{abstract}


\end{frontmatter}

\section{Introduction}

The emergence of transformer-based large language models (LLMs) has fundamentally transformed natural language processing capabilities, enabling sophisticated text understanding, analysis, and generation tasks across diverse domains \cite{vaswani2017attention}. These advances have catalyzed the development of AI-driven solutions in sectors ranging from healthcare and education to automated documentation and diagnostic support \cite{MENG2024109713, Baierl2023}. However, to make use of LLMs in software, we need to understand the unique circumstances in different domains since general-purpose chatbots are reluctant to solve domain-specific problems well~\citep{bespokeLegal,liang2024encouragingdivergentthinkinglarge}. The use of different types of LLM solutions in financial report analysis represents a relatively underexplored frontier with significant potential for workflow automation and decision-making enhancement \cite{alnaqbi2024enhancing, Yu2024}.

Financial report analysis constitutes a critical function for businesses, investors, and regulatory bodies, yet traditional approaches remain predominantly manual, time-intensive, and susceptible to inconsistencies \cite{Akcan2025}. Corporate finance workflows involve substantial repetitive tasks including data analysis across multiple formats, financial statement preparation, and extensive fact-checking procedures. While recent NLP advancements demonstrate promising capabilities for automating financial document processing and insight extraction \cite{Wu2023}, substantial knowledge gaps persist regarding the practical integration of LLMs into real-world financial systems. Key challenges encompass data sensitivity requirements \cite{privacy_preserving}, domain-specific terminology complexities \cite{araci2019finbertfinancialsentimentanalysis}, and critical demands for interpretability and accuracy in financial contexts.

Despite the recognized potential of LLMs for financial automation, existing research provides limited guidance on the software engineering practices necessary for successful implementation. This study addresses this gap by examining the technical and organizational challenges inherent in integrating LLMs into financial workflows, aligning with Uchitel et al.'s \cite{uchitel2024scoping} framework for AI-enabled systems engineering. Through an exploratory case study conducted in collaboration with Getinge AB, this research investigates the software engineering complexities of making use of open-source and commercial LLM-based financial report analysis systems.

The study is guided by two primary research questions: \textbf{RQ1:} \textit{What are the primary software engineering challenges in integrating LLMs into domain-specific report analysis workflows?} and \textbf{RQ2:} \textit{How do open-source and commercial LLMs compare in terms of software engineering complexity, performance, and maintainability?}. The investigation focuses specifically on system architecture design, workflow optimization, and reliability evaluation within financial reporting contexts, contributing to both AI systems engineering and practical financial technology implementation.

This paper proceeds as follows: Section 2 presents an overview of relevant literature on LLM integration and financial automation systems. Section 3 reviews related work in more detail, Section 4 outlines the research methodology and case study design, and Section 5 presents the results. Section 6 discusses the result regarding software engineering challenges and implementation trade-offs together with threats to validity while Section 7 concludes the paper and suggests future research directions.

\section{Background}
This section provides the theoretical foundation for the study by introducing key approaches to open-source and commercial LLM solutions and more detail of our selected case of financial analysis. 

\subsection{Natural Language Processing and Large Language Models}

Natural Language Processing is a subfield of artificial intelligence focused on enabling systems to understand and generate human text through machine learning, deep learning, and linguistic rules \cite{jm3}. The field evolved from early rule-based systems like ELIZA \cite{Weizenbaum1966} through statistical methods utilizing probabilistic models such as Hidden Markov Models and n-gram models \cite{chen1999empirical}, to modern deep learning approaches featuring neural architectures like RNNs and LSTMs \cite{hochreiter1997long}.

The introduction of the Transformer architecture \cite{vaswani2017attention} marked a paradigm shift, enabling parallel processing and improved modeling of long-range dependencies through self-attention mechanisms. The self-attention mechanism calculates relationships between tokens using:
\[
\text{Attention}(Q, K, V) = \text{softmax} \left( \frac{QK^T}{\sqrt{d_k}} \right) V
\]
where Query (Q), Key (K), and Value (V) are vectors derived from input embeddings. Multi-head attention performs multiple simultaneous calculations to capture various token relationships \cite{vaswani2017attention}.

Transformers consist of encoder and decoder components. Encoder-only models like BERT \cite{devlin2018bert} excel at text analysis tasks, decoder-only models like GPT \cite{radford2018improving} specialize in text generation, while encoder-decoder models like T5 handle text-to-text transformations. Modern LLMs undergo pretraining on large-scale datasets followed by fine-tuning for specific domains \cite{brown2020language}.

\paragraph{Prompting and Agent-Based Approaches}
Effective prompt design significantly impacts LLM output quality. Well-constructed prompts provide contextual information, clear instructions, and concrete examples \cite{anthropic2024promptguide}. Key techniques include Chain-of-Thought prompting, which instructs models to show intermediate reasoning steps \cite{wei2022chain}, and Few-Shot prompting, which provides task examples within the prompt \cite{brown2020language}. Modern models support extensive context windows (up to 200,000 tokens in Claude 3.7 Sonnet \cite{anthropic2025claude37} and 1 million in Gemini 2.5 \cite{google2025gemini25pro}), though information placement affects utilization effectiveness \cite{liu2024lost}.

LLMs can function as autonomous agents performing reasoning and decision-making tasks \cite{peng2023agentbench}. The ReAct framework \cite{yao2023react} demonstrates interleaving reasoning with tool use and environment feedback. Common orchestration patterns include prompt chaining (sequential task decomposition), parallelization (concurrent sub-task execution), orchestrator workflows (centralized task delegation), routing (task classification and assignment), and evaluator-optimizer loops (iterative refinement) \cite{anthropic2024agents}. However, multi-agent systems face challenges including specification issues, inter-agent misalignment, and task verification failures \cite{pan2025multiagent}.

Recent reasoning models like OpenAI's o3 and Anthropic's Claude 3.7 Sonnet internalize these techniques through reinforcement learning on chain-of-thought trajectories and extended thinking modes, enabling complex multi-step workflows through simple natural language prompts \cite{jaech2024openai, anthropic2025claude37card}.

\subsection{Implementation Challenges}
LLM integration presents significant engineering challenges. High computational requirements necessitate powerful hardware, with models like GPT-4o containing over a trillion parameters \cite{shahriar2024putting}. Optimization techniques including quantization (reducing weight precision) and pruning (removing less important parameters) help reduce computational demands \cite{frantar2022gptq, ma2023llm}.

Data security and privacy concerns arise from potential unintentional learning of sensitive information and vulnerability to prompt-based attacks, model inversion, and membership inference attacks \cite{Usenixsec}. Robust security protocols including differential privacy, encryption, and secure API gateways are essential \cite{feretzakis2024trustworthy}.

LLMs are prone to hallucinations, generating plausible but incorrect information due to their probabilistic nature \cite{Bender2021}. General-purpose models often lack contextual understanding in specialized domains, with studies showing that domain-specific models provide more accurate interpretations than generalized alternatives \cite{Hajikhani2024, Ahmed2024}. These limitations necessitate additional validation mechanisms for AI-generated financial analyses.

\subsection{Financial Statement Analysis and Reporting}
Financial statements provide a snapshot of a company's financial condition, offering insights into performance, profitability, liquidity, and overall economic position \cite{berk2013fundamentals}. The primary statements include income statements (presenting revenue, expenses, and profits with details on operational efficiency), balance sheets (detailing assets, liabilities, and shareholder equity), and cash flow statements (providing information about cash inflows and outflows across operating, investing, and financing activities) \cite{berk2013fundamentals}.

Traditional financial analysis relied heavily on rule-based systems including expert systems that emulate human decision-making through IF-THEN logic \cite{Expert_system}, spreadsheet analysis using tools like Excel for data manipulation and ratio calculations \cite{stanfordCS102}, and various analytical techniques such as ratio analysis and vertical/horizontal analysis for examining financial relationships and trends \cite{berk2013fundamentals}. However, these approaches struggle with modern financial documents' complexity, particularly when processing unstructured data such as management commentaries and footnotes \cite{lewis2019fad}. Machine Learning and NLP solutions offer automated parsing and analysis of both textual and numerical data at scale, though challenges persist including inconsistent data formats, domain-specific tuning requirements, and information volume management.

\section{Related Work}
The following parts of this section review literature relevant to the use of LLMs in software engineering and the current research in LLM for the financial context. 

\subsection{Software Engineering for LLM-based systems}
AI and machine learning have become increasingly integrated in software engineering, both as tools to improve development workflows and as components within software systems \cite{shafiq2020machine, martinez2022software}. AI-driven techniques are, for example, used in automated testing, code generation, bug detection, software maintenance, improving software quality and developer productivity. However, as AI components become more deeply embedded in software products, they introduce new challenges related to system reliability, explainability, and risk management.

\paragraph{Software Engineering for AI}
To better understand the role of AI in software engineering, efforts have been made to categorize AI applications. One such attempt is the AI-SEAL taxonomy by Feldt et al. \cite{feldt2018ai}, which classifies AI use based on point of application (development process, product, or runtime), type of AI, and level of automation. Their work highlights that AI integration at the runtime level introduces higher risks, requiring robust governance mechanisms to ensure system reliability and compliance.

Within the broader AI landscape, LLMs have emerged as a distinct category, primarily used to support software development tasks such as code generation, documentation, and debugging. However, research on LLMs as components within software systems remains limited \cite{weber2024llm}.  The authors further underscore that there is a need for new software engineering practices. Specifically, it calls for treating prompts as software assets and developing robust testing and monitoring frameworks for systems that rely on LLM-generated outputs. Although studies on LLM-integrated applications from a software engineering perspective remain limited, the field is rapidly emerging.

\paragraph{Software Engineering for LLM-based applications }
Chen et al. \cite{CHEN2024Framework} propose a feature requirement framework for developing applications based on LLMs in industry to address challenges with implementing LLMs in an industry setting. Feature requirements need to be addressed because of the limitations of current LLMs, coming from hallucination, the black box nature, and weak specialized knowledge. 

Studies concerning the architecture of LLM-based systems are a growing area of research. Lu et al \cite{lu2024taxonomy} propose a taxonomy of foundation model-based systems that outlines key architectural considerations across three categories. First, the pre-training and adaptation dimension distinguishes approaches such as in-context learning, fine-tuning, and distillation, each carrying trade-offs in control, accuracy, and infrastructure requirements. Second, the system architecture design dimension introduces the idea of foundation models as architectural connectors that serve roles such as communication, coordination, and conversion across system components. This framing positions LLMs as middleware capable of orchestrating complex, multi-step workflows. Third,  Responsible AI by design dimension presents mechanisms for managing risk and ensuring compliance, including verifiers and prompt guardrails. An insight from their work is that in-context prompting alone is often insufficient for tasks requiring high output fidelity. Architectural patterns such as prompt chaining, task decomposition, or verifier modules are necessary, where interpretability, traceability, and formatting consistency are essential. The taxonomy also draws attention to deployment trade-offs between external and sovereign models, a distinction that is especially relevant in data-sensitive and regulated environments \cite{lu2024taxonomy}.

Expanding on the architectural considerations, Mahr et al. \cite{mahr2024reference} propose a reference architecture for deploying LLM applications in constrained environments, with a focus on addressing data security, limited connectivity, and infrastructure constraints. The architecture defines five stages: data preparation, model strategy selection, model evaluation, deployment, and prompt engineering. It outlines deployment options suitable for environments with strict data protection requirements, including on-premises and edge-based solutions. The component of the model strategy includes alternatives such as parameter-efficient fine-tuning (PEFT), full fine-tuning, and retrieval-augmented generation (RAG), each with different trade-offs in resource usage and control. The paper also incorporates LLMOps practices to support ongoing evaluation and adaptation, which are relevant for maintaining reliability and compliance in regulated domains.

Wang et al. \cite{wang2024survey} highlight key architectural and operational aspects of LLM-based autonomous agents. They propose a modular framework where LLMs are enhanced with memory, planning, and action modules, enabling them to handle complex workflows, recall context, and generate structured outputs. The study emphasizes that agents can acquire expertise through fine-tuning or prompt-based adaptation. In addition, they stress the importance of rigorous evaluation, combining objective benchmarks with subjective expert assessments to ensure reliability.
 
Research has also explored practical frameworks for developing LLM-based applications. One notable example is LangChain \cite{topsakal2023llm}, an open-source library designed to facilitate rapid development of LLM-based applications. LangChain provides modular abstractions for interacting with various data sources and applications, enabling developers to build flexible and scalable LLM-driven solutions. Other examples of frameworks are Hugging Face’s SmolAI Agents \cite{huggingface_smolagents} designed for lightweight autonomous agents when computational resources are a constraint and AutoGen \cite{wu2023autogen} facilitates multi-agent collaboration. These frameworks contribute to the growing ecosystem of LLM development and agent coordination 

\paragraph{Engineering challenges}
Regarding the challenges that developers face in real-world settings, a study that conducted 26 interviews with software engineers \cite{parnin2023copilot} identified four key issues. First, interacting with LLMs demands extensive prompt engineering, output modification, workflow development, and unpredictability management. Second, testing and validation remain challenging due to the absence of standardized metrics and the necessity for custom testing solutions. Third, ensuring safety, privacy, and compliance introduces concerns related to user consent and regulatory adherence in AI-driven actions. Finally, developers struggle with learning and tool integration due to the lack of centralized resources and the continuously evolving nature of workflows. 

\subsection{Current Research on Large Language Models in Finance}
The capabilities of LLMs have enabled their application in various financial analysis tasks, including text summarization, information extraction, sentiment analysis, and financial forecasting \cite{lee2024survey}. The versatility of LLMs for different tasks has led to the development of specialized models and architectures to address domain-specific challenges. Ongoing research into techniques like prompt engineering and multi-agent frameworks aim to enhance their performance in financial systems further.
 
\paragraph{LLMs for Financial Tasks}
Several LLMs have been developed for financial tasks. One of the first examples is FinBert \cite{huang2023finbert}, which is built on Google's BERT architecture \cite{devlin2018bert}. These are a set of models that are built using only financial data or combined with general domain data.  FinBert is designed for structured NLP tasks such as sentiment analysis, information extraction, and classification. Another notable example is BloombergGPT \cite{wu2023bloomberggpt}, pre-trained on a massive set of financial data combined with general knowledge, which has demonstrated the potential for text generation, question and answering, text summarization, and financial forecasting within the domain.

Additionally, research has explored models through instruction tuning for financial tasks. Instruction tuning involves training models on financial datasets with instruction sets for specific tasks \cite{zhang2023instruction}. For example, Zhang et al. \cite{Zhang2023} demonstrated that instruction tuning can significantly enhance LLMs for financial sentiment analysis. Their model, Instruct-FinGPT, based on LLaMa, outperformed FinBERT and GPT-3.5 across multiple datasets in terms of accuracy and F1 score. Advancements included improved numerical sensitivity, enabling better sentiment interpretation of financial indicators, and enhanced contextual understanding, which facilitated accurate sentiment prediction in nuanced scenarios.
Other notable examples using this approach include FinGBT \cite{yang2023fingpt} and PIXIU \cite{xie2023pixiu}.

\paragraph{Enhancing LLMs with Prompt Techniques}
Zhang et al. \cite{Zhang2023} demonstrate that instruction tuning combined with Retrieval-Augmented Generation (RAG) significantly enhances LLM performance in financial sentiment analysis. RAG provides the model with external contextual data, thereby addressing some of the LLM's limitations in understanding complex financial language. This approach achieves notable improvements, with accuracy and gains in the F1 score of 15–48\%. 

The rapid development of the capabilities of general-purpose models gives them the potential to perform effectively within the financial domain. This potential is further amplified when combined with prompt engineering techniques such as Chain of Thought (CoT), which have proven effective in guiding LLMs toward more accurate and transparent reasoning processes. The effectiveness of this was demonstrated in \cite{Kim2024}. Their study compared human analysts with GPT-4 \cite{openai2023gpt4}, a general-purpose model, to predict future earnings using two approaches: simple prompting and CoT prompting. Without CoT, GPT-4 achieved an accuracy of 52.33\%, marginally better than human analysts at 52.71\%. However, with CoT prompting, the accuracy of GPT-4 increased significantly to 61.45\%, outperforming human analysts. Notably, GPT-4 achieved this without access to the narrative context typically available to human analysts, highlighting the potential of CoT to improve LLM reasoning capabilities.

Research has also focused on evaluating LLM summarization capabilities in the financial domain. Yang et al. \cite{yang2024evaluating} conducted a comprehensive benchmark comparing GLM-4, Mistral-NeMo, and LLaMA 3.1 (70B variants) on financial report summarization tasks. Their evaluation combined traditional metrics (ROUGE, BERTScore) with an LLM-based qualitative review, assessing dimensions such as accuracy, informativeness, and coherence. The results showed that while models like LLaMA 3.1 achieved higher accuracy and coherence scores, they incurred significantly longer processing times. The study also found that model size alone does not guarantee better performance, highlighting the importance of prompt design, domain fit, and system constraints.  

\paragraph{Architectural Innovations for Financial LLM Systems}
To enhance the capabilities of systems utilizing LLMs in the financial context, research is also exploring the development of architectures, such as multi-agent frameworks. Mixture of experts \cite{masoudnia2014mixture} is one approach, where individual agents function as specialized experts, handling their distinct tasks to enhance reasoning, generalization, and task-specific performance \cite{Cai2024}. This design allows for increased model capacities without the need for increased computational demand. This is achievable by only activating the relevant expert during inference.

Yu and Tiwari \cite{Yu2024} propose FinTeamExperts, a framework using a Mixture of Experts (MoE) approach with large language models (LLMs) customized for financial analysis. It employs three specialized models that simulate distinct roles: Macro Analysts, Micro Analysts, and Quantitative Analysts, each trained on domain-specific data. Through a two-phase training process, pre-training on role-specific corpora, and instruction-tuning with routing gates, therefore integrating diverse expertise for financial tasks. Upon evaluation on public datasets, the framework outperformed comparable and larger models.

\paragraph{Integration of LLMs in the financial domain}
Tavasoli et al. \cite{tavasoli2025responsible} present a framework that covers some core challenges financial institutions face when deploying LLMs. Their takeaway is that successful and responsible adoption of LLMs in finance requires navigating several technical, regulatory, and strategic factors. Through a six-part framework, the authors outline sequential decisions institutions must make: 1 determining whether an LLM is necessary over simpler methods, 2 establishing robust data governance under regulations such as GDPR, 3 implementing targeted risk management practices including explainability, human-in-the-loop oversight, and adversarial testing, 4 embedding ethical oversight to ensure fairness and transparency, 5 assessing the return on investment from both tangible and intangible perspectives, and 6 choosing an appropriate implementation strategy \cite{tavasoli2025responsible}. 

The authors further discuss the tradeoffs that different implementation strategies imply.  For example, open-source models offer greater transparency and flexibility but require significant internal resources and expertise. In contrast, proprietary models provide stronger performance out of the box at the cost of limited control and potential compliance risks. Similarly, in-house deployment supports strict data control but demands high technical capacity, whereas vendor-managed solutions ease operational burden but raise concerns over data exposure and vendor lock-in. In terms of model adaptation, the authors argue that prompt engineering is an agile and easy method but lacks reliability, full fine-tuning offers better accuracy at the cost of a high computational burden, and RAG improves factual accuracy through external references, however, with increased infrastructure complexity \cite{tavasoli2025responsible}.

\subsection{Summary and gaps}
Software engineering research has begun to explore architectural frameworks, development practices, and operational challenges for LLM-based applications. While this work provides valuable insights, much of it remains general-purpose, is mainly theoretical, and lacks application-specific depth in domains like finance.

In parallel, existing research on LLMs in the financial domain has primarily focused on enhancing model performance for specific tasks such as sentiment analysis, summarization, and forecasting. Techniques like instruction tuning, domain-specific pretraining, prompt engineering, and architectures have been explored to enhance capabilities. However, these efforts are often evaluated in isolation and do not extend to a real integration of LLMs into financial workflows.

As a result, a gap exists at the intersection of LLMs' capabilities and system-level implementation. Specifically, there is limited research on how different LLMs can be designed into financial environments. This study addresses this gap by investigating the practical challenges of developing and deploying open-source or commercial LLM-driven artifacts in financial reporting workflows to inform future system design and implementation practices for similar reporting contexts.

\section{Method}
In relation to the ABC framework \cite{stol2018abc}, the study is conducted as a field study, which means the research is taking place in a real-world setting without actively controlling or altering the environment. This approach facilitates a nuanced understanding of the challenges and opportunities associated with integrating LLMs into domain-specific workflows. However, it may come at the expense of precision and generalization compared to studies conducted in a more controlled environment. Within the context of field studies, this thesis could be categorized as an exploratory field study following the definition by Höst et al. \cite{runeson2012casestudy}. The goal is to generate new insights, identify key challenges, and propose hypotheses for future exploration. To achieve this, the study applies Design Science Research (DSR) \cite{hevner2004design}, which blends practical problem-solving with knowledge generation through iterative design and evaluation.

\subsection{Methodology}
The study follows the Design Science Research (DSR) framework proposed by Pfeffers et al. \cite{Peffers2007}. Design Science Research is particularly suitable for research that aims to develop, implement, and evaluate artifacts, making it a suitable approach to explore how different types of LLMs can be utilized within domain-specific analysis and reporting. Unlike traditional empirical research, which seeks to describe and explain phenomena, DSR focuses on solving problems by designing and evaluating solutions in real-world environments \cite{Peffers2007}. Given that this thesis aims to develop and evaluate an AI-driven solution to identify its challenges, DSR provides a structured process to guide the design, implementation, and validation of the proposed system. 

\begin{figure}[H]
    \centering
    \includegraphics[width=1\linewidth]{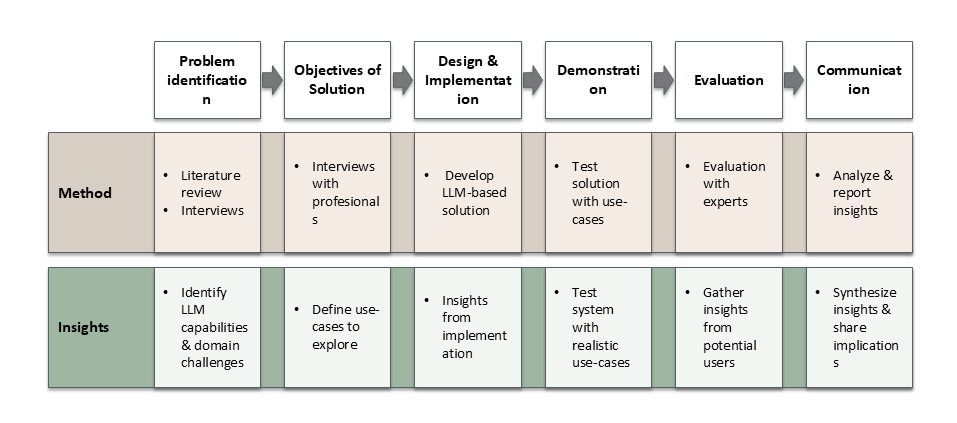}
    \caption{Alignment of DSR steps with the specific methods applied and the insights each step is intended to generate.}
    \label{DSROverview}
\end{figure}

An overview of the research methodology is illustrated in \ref{DSROverview}, which maps each methodological step to the design science research process while highlighting the practical insights gained during the phase.

\subsection{Research Setting}
This study is conducted in collaboration with the finance department at Getinge AB, a Swedish MedTech company specializing in healthcare and life sciences solutions. Getinge provides a wide range of products and services and is publicly listed on the Nasdaq Stockholm stock exchange. The company operates globally, with a presence in over 40 countries, focusing on areas such as intensive care, operating rooms, sterile reprocessing, and infection control.

The finance department at Getinge operates within a sensitive financial environment, handling confidential data that demands a high level of accuracy and reliability in reporting and analysis, making it a valid case from a more safety-critical domain.  

\subsection{Literature review}
The first part of the study involved conducting a literature review, following the guidelines proposed by Keele \cite{keele2007guidelines}. The review was conducted in three phases: planning, conducting, and synthesizing the results. The review aimed to gain an understanding of the current research on software engineering for LLM-based systems and LLMs within the financial domain. 


The review considered common software engineering challenges encountered when integrating AI-driven systems into industry settings. These themes helped structure the literature synthesis and provided a foundation for understanding the broader context of LLM integration and the specific context of financial workflows.


The search was conducted in IEEE Xplore, ACM Digital Library, ScienceDirect, and Google Scholar. To the extent possible, peer-reviewed articles were prioritized. However, given the rapid advances in research related to the field, many recent studies have not yet undergone peer review. Additionally, a substantial portion of relevant research originates from industry. To ensure the most up-to-date information, non-peer-reviewed papers and online articles were also considered.

\subsection{Overview of Research Activities}
The research activities were structured into a series of exploratory and design-oriented phases, which are referred to as cycles in this study. Following the methodological framing and literature review, Cycle 0 explored the problem space through engagement with domain experts, aiming to identify relevant use cases and uncover potential challenges associated with designing LLM-based solutions into financial workflows. Based on these insights, two design cycles were conducted to implement a solution. Finally, an evaluation was conducted together with experts in the domain to asses and compare the solutions. 

\paragraph{Cycle 0: Domain Exploration}
Cycle 0 served as the exploratory phase of the study, aimed at gaining a contextual understanding of the department and its operations. This was achieved through two focus group sessions followed by a task analysis. The focus groups were conducted with business controllers to uncover high-effort cognitive tasks and understand current practices in generating narrative financial commentary.

The insights gained from this cycle informed the formulation of system objectives and design considerations for the LLM-based artifacts developed in Cycles 1 and 2.

An overview of the conducted activities and participating stakeholders is presented in Table~\ref{tab:prestudy} below. A more detailed account of this process is provided in the subsequent section.

\begin{table}[ht]
    \centering
    \begin{tabularx}{\textwidth}{@{}Xlll@{}}
        \toprule
        \textbf{Activity}             & \textbf{Participants}         & \textbf{Duration} & \textbf{Location} \\
        \midrule
        Focus Group                   & 3× Business Controllers       & 1 hour            & On-site           \\
        Focus Group                   & 3× Business Controllers       & 1 hour            & On-site           \\
        Workflow Walkthrough          & 2× Business Controllers       & 1 hour            & On-site           \\
        \bottomrule
    \end{tabularx}
    \caption{Overview of exploratory study activities}
    \label{tab:prestudy}
\end{table}

\paragraph{Cycle 1 and 2 Design and Development}
Following the initial exploration, two design and experimentation cycles were carried out to investigate the applicability of two different LLM implementation strategies for automating tasks in the financial analysis domain.

\textbf{Cycle 1:} focuses on developing and testing a locally hosted solution using lightweight open-source models. The procedure involved implementing a modular, multi-agent workflow where agents processed structured financial data and generated summarizing commentary. 

\textbf{Cycle 2:} explores the use of a commercially available LLM (OpenAI’s GPT-4o) accessed via API. The procedure mirrored that of Cycle 1 in terms of workflow and summarization tasks, but prioritized the evaluation of output quality, interpretability, and ease of integration. 

Together, these cycles provided insights into different approaches to LLM design in real-world reporting contexts by applying the solutions to the finance reporting context.

\paragraph{Evaluation}
To evaluate the outputs from both artifacts, AI-generated financial summaries were compiled and shared with domain experts working in financial reporting. The initial plan was to conduct a live, structured evaluation session in which the summaries would be assessed collaboratively. However, this format proved impractical, since we wanted to assess multiple aspects such as the content's clarity, accuracy, language used, and potential integration into existing workflows. 

The evaluation format was therefore adapted, participants were given access to the summaries and a set of guiding questions, which they could review independently at their own pace over the course of a week. A follow-up session was then held to discuss their reflections, during which follow-up questions and requests for clarification could be addressed.

\begin{table}[ht]
    \centering
    \begin{tabular}{@{}lllc@{}}
        \toprule
        \textbf{Activity}                   & \textbf{Participants}               & \textbf{Duration} & \textbf{Format} \\
        \midrule
        Initial review  & 2× Business Controllers             & 1 hour            & On-site         \\
        Independent review of summaries    & 2× Business Controllers             & N/A               & N/A             \\
        Follow-up discussion session       & 2× Business Controllers             & 1 hour            & Online          \\
        \bottomrule
    \end{tabular}
    \caption{Overview of evaluation activities}
    \label{tab:evaluation_overview}
\end{table}

To guide the evaluation, a set of open-ended questions is grouped under six key themes. These themes focused on areas such as language clarity, content accuracy, perceived usefulness, integration into participants' workflows, and trust in AI-generated content. The themes and associated guiding questions can be found in the Appendix \ref{tab:evaluation_questions}.

\section{Design Iterations and Evaluation}
The following section presents the design, implementation, and evaluation of a solution designed to automate financial commentary using LLMs. It begins by outlining the exploratory work conducted to identify use cases for LLM-based automation within the financial reporting process. This is followed by a description of the selected use cases and observations made.

Subsequent sections detail the development of two prototypical solutions, first a local LLM-based workflow and a cloud-based alternative using GPT-4o. This is followed by observations made from experimentation.

The section concludes with an evaluation of both prototypes, highlighting findings across critical dimensions such as, accuracy, usability, and trustworthiness.

\subsection{Cycle 0: Domain Exploration}
To identify the relevant areas for LLM-based automation within the financial department of Getinge AB, two exploratory meetings were conducted with representatives of the business control team. The initial session introduced the thesis scope and aimed to establish a foundational understanding of the department's operations.

The discussion then focused on identifying processes characterized by high cognitive workload and frequent manual interpretation. These were deemed particularly suitable for LLM integration. Through these dialogues, the task of preparing financial statements was identified as a promising entry point. Other areas of interest included synthesizing insights from unstructured information sources and performing financial forecasting.

To deepen the understanding of reporting workflows, a follow-up session was held to map the steps involved in generating narrative commentary for financial statements. Based on these discussions, three specific use cases were defined as suitable for initial implementation and evaluation.

These use cases represent scenarios that were both considered contextually relevant and technically feasible within the scope of the study. They serve as concrete points of departure for exploring the practical challenges of designing LLM solutions in the financial reporting context. The selected tasks reflect typical areas of manual analysis: identifying deviations in financial metrics, summarizing key drivers behind those deviations, and producing structured qualitative commentary.

\paragraph{Objectives of the Solution}
To clarify what an LLM-based solution should achieve, the reporting process was analyzed in collaboration with two business controllers. The aim was to identify the underlying data sources, determine the relevant information, and clarify the cognitive steps involved in the process. Emphasis was placed on capturing the thought process typically employed by business controllers when formulating these summaries, as this reasoning is what the LLM-based solution is intended to emulate and support.

The process begins by detecting significant changes in key metrics, such as order intake or net sales, for specific product segments. These changes are then traced to underlying drivers by examining performance across product lines and geographic regions. Finally, business controllers assess the relative importance of these drivers and distill the findings into concise narrative summaries for inclusion in financial reports.

The goal of the LLM-based system is to emulate and support this workflow by automating the identification, synthesis, and articulation of key insights. This involves mimicking not just the data processing, but also the judgment applied in prioritizing and articulating findings.

\paragraph{Financial Use Cases for LLM-Based Automation Testing}
The following is a description of the workflow that the LLM is intended to facilitate, outlining its objectives, key insights, process, and evaluation criteria.

\textbf{UC1: Generating Summarizing Comments for Income Statements}  

\textbf{Objective:} Automatically generate qualitative comments on the income statement presented alongside the financial statement for a given period.  

\begin{itemize}
    \item \textbf{Key Insights:} Identify increases or decreases in profits, OPEX, and business area-specific trends.
    \item \textbf{Process:} The data is analyzed and compiled into a CSV file, highlighting deltas of changes in Revenue and Costs. Based on this data, the LLM model should infer key highlights and drivers of changes during the period.
    \item \textbf{Evaluation:} Compare AI-generated insights against human analysis.
    \item \textbf{Example:} Cost related to [category] is the main driver for increases in OPEX.
\end{itemize}

\textbf{UC2: Summarizing Comments in Changes in Order Intake}  

\textbf{Objective:} Generate a summary of increase and decrease in order-intake divided into business-area, product-area and find the drivers within the product-lines. 

\begin{itemize}
    \item \textbf{Key Insights:} Highlight notable drivers for the changes in order-intake between two periods in a qualitative summary. For example, the current and previous months. 
    \item \textbf{Data Source:} Structured order intake data containing numerical figures for two periods, along with categorical information such as business areas, product lines, and regions, extracted either from a database or provided in a CSV format.
    \item \textbf{Evaluation:} Compare AI-generated summaries to human-generated insights.
    \item \textbf{Example Output:}  [Product Line] in [Region] as main growth driver. [Product Line] up in all regions. [Product Line] increasing, mainly in [Region]. Decrease from [Product Line] party products in [Region].[Product Line] decreasing in [Region].
\end{itemize}

\textbf{UC3: Summarizing Comments in Changes in Net Sales}

\textbf{Objective:} Generate a summary of increases and decreases in net sales, divided into business areas and product areas, and identify the key drivers within the product lines.
\begin{itemize} 
\item \textbf{Key Insights:} Highlight notable drivers for the changes in net sales between two periods in a qualitative summary, for example, between the current and previous month. 
\item \textbf{Data Source:} Structured net sales data containing numerical figures for two periods, along with categorical information such as business areas, product lines, and regions, extracted either from a database or provided in a CSV format. 
\item \textbf{Evaluation:} Compare AI-generated summaries to human-generated insights. 
\item \textbf{Example Output:} [Product Line] in [Region] as main growth driver. [Product Line] up in all regions. [Product Line] increasing, mainly in [Region]. Decrease from [Product Line] party products in [Region]. [Product Line] decreasing in [Region]. \end{itemize}

\paragraph{Challenges with integrating LLMs in the financial context}
A practical challenge observed early during the exploration of applying LLMs to financial analysts workflows was that the data used in their work comes in varied formats, requiring substantial data preprocessing. Financial information used in analysis and reports is collected from multiple sources and formats, including Excel spreadsheets, PowerPoint presentations, ERP systems, and oral communication during meetings. These inconsistencies in data structure and terminology require some form of preprocessing to ensure that the input to the LLM is usable and appropriate for the task given to the model.

It became evident during this phase that any practical application of LLMs would require tailored preprocessing logic. Although the use cases later tested in the study relied on structured data and custom scripts to extract and format this data into prompt-ready text, this approach was closely tied to the structure and content of that particular case. The early exploration suggested that similar preprocessing efforts would likely be necessary for other types of reports or financial analyses, each requiring its own logic to handle differences in layout, terminology, and contextual focus.

An obstacle when designing LLMs into financial analysis workflows is the sensitive nature of the data involved. Rules that the department needs to adhere to prohibit sharing non-public financial information to third parties, as in the case of LLMs hosted by a third party. As a result, experiments involving these models had to rely on synthetic or anonymized data to ensure compliance. Importantly, this limitation is not unique to the research setting, it reflects the operational reality for employees as well, who are restricted from sharing sensitive information externally due to regulatory requirements and internal governance policies.

These constraints were the reason Use Case 1 was not pursued in the study, as the data used for the task comes in varied data formats, and making a realistic transformation on the data would be infeasible without making the data non-realistic. Instead, use cases 2 and 3 were continued, as they were more practical to implement under the given constraints.

\subsection{Cycle 1: Local setup}
In response to the challenges identified in the initial exploration, particularly on restrictions on sharing sensitive data externally, the first cycle focused on testing a workflow using local LLMs. This would support data privacy by keeping all processing on-site and is potentially lightweight enough to operate on standard office hardware, a consideration given the departments lack of dedicated infrastructure. It also presents a cost-effective alternative to commercial solutions. Furthermore, if the tasks assigned to the system can be handled effectively by a simpler mode, there is little justification for adopting a more complex solution, especially one that is more computationally demanding, costly, and introduces additional challenges concerning deployment.

Use Cases 2 and 3, which focus on summarizing changes in order intake and net sales, respectively, across business and product areas, were selected for continued exploration. Both relied on structured datasets that could be anonymized or synthetically replicated, making them suitable for local experimentation while respecting data sensitivity constraints.

\paragraph{Data pre-processing}
Before either solution could generate summaries, it was essential to extract and structure the relevant data in a format that the language models could interpret effectively. The raw data was initially provided as CSV spreadsheets containing thousands of entries. To prepare this data for model input, we used Python and the Pandas library to carry out pre-processing steps, including calculating the differences in net sales or order intake between two periods, grouped by product line and region.
To ensure the input was both manageable and meaningful, the dataset was then condensed into a focused and structured summary of key trends.

While Excel files were used in this prototype for convenience and accessibility, a real-world implementation would likely rely on direct integration with a database or business intelligence system. Table~\ref{tab:product_differences_modified} shows an example of the data fed into the model for interference.  

\begin{table}[H]
    \centering
    \caption{Modified Example Data as input to model.}
    \label{tab:product_differences_modified}
    \begin{tabular}{l l r r}
        \hline
        \textbf{Product Line} & \textbf{Region} & \textbf{Total Difference} & \textbf{Contribution (\%)}\\
        \hline 
        CCVE & EMEA - EMEA & -5,619,037.0 & -5.48 \\
        CCVE & APAC - Asia/Pacific & -17,886,827.0 & -16.49 \\
        CCVE & AMER - Americas & 89,963,264.0 & 65.35 \\
        CCSE & EMEA - EMEA & 32,582,792.0 & 34.23 \\
        CCSE & APAC - Asia/Pacific & 18,493,137.0 & 17.90 \\
        CCSE & AMER - Americas & 7,142,562.0 & 6.85 \\
        CCOT & APAC - Asia/Pacific & 654,822.0 & 0.49 \\
        CCOT & EMEA - EMEA & -2,874,312.0 & -3.15 \\
        CCOT & AMER - Americas & -12,589,374.0 & -11.78 \\
        CCHH & EMEA - EMEA & 3,405,672.0 & 2.78 \\
        CCHH & APAC - Asia/Pacific & 2,598,442.0 & 2.55 \\
        CCHH & AMER - Americas & -1,078,936.0 & -1.02 \\
        CCHD & EMEA - EMEA & -2,874,098.0 & -2.71 \\
        CCHD & APAC - Asia/Pacific & -6,945,325.0 & -6.51 \\
        CCHD & AMER - Americas & -785,214.0 & -0.73 \\
        CCAA & APAC - Asia/Pacific & -1,678,432.0 & -1.35 \\
        CCAA & EMEA - EMEA & 11,230,874.0 & 9.99 \\
        CCAA & AMER - Americas & 5,093,287.0 & 5.15 \\
        \hline
    \end{tabular}
\end{table}

It lists the product lines and their corresponding regions, along with the aggregated difference in order intake or net sales between two periods ("Total Difference"). The column 'Contribution (\%)' indicates the relative impact of each product line and region on the overall change, guiding the model on the importance of individual drivers.

\paragraph{Local LLM experimentation}
To facilitate experimentation with models deployed on local machines, we used Ollama \cite{ollama}, a tool that provides a setup for running and interacting with various open-source LLMs. Initially, two smaller models, LLaMA3.2 and Qwen2.5, were tested, each with approximately 3 billion parameters, to explore their suitability. As experimentation progressed, we moved on to models with around 8 billion parameters. However, inference at that scale became significantly slower due to hardware limitations, as the models were running solely on CPU without GPU acceleration. A summary of the models tested in the experimentation can be found in Table \ref{tab:tested_models}
\begin{table}[ht]
    \centering
    \begin{tabular}{@{}llllll@{}}
        \toprule
        \textbf{Model} & \textbf{Size} & \textbf{Reasoning} & \textbf{Quantized} & \textbf{Context Window} & \textbf{Type} \\
        \midrule
        Qwen 2.5        & 3B  & No   & 4-bit & 32k  & Instruct \\
        LLaMA 3.2       & 3B  & No   & 4-bit & 128k & Instruct \\
        Qwen 2.5        & 7B  & No   & 4-bit & 32k  & Instruct \\
        LLaMA 3.1       & 8B  & No   & 4-bit & 128k & Instruct \\
        DeepSeek-R1     & 8B  & Yes  & 4-bit & 128k & Distilled \\
        \bottomrule
    \end{tabular}
    \caption{Overview of models tested in early stages.}
    \label{tab:tested_models}
\end{table}

During early testing, single-prompt inputs were explored, in which instructions and data were provided to the model within a single input. This approach was found to be ineffective, leading to hallucinations, poor structural coherence, and factual inaccuracies. As a result, a multi-agent workflow was employed. In this design, distinct roles were assigned as specific agents, each tasked with handling a specific sub-task of the overall task. LLaMa3.1 was used in the later setups, since after the initial testing, it proved to be the most stable for the given task.

\paragraph{Implementation of Multi-Agent Workflow}
To implement the four-agent design, a Python-based pipeline was developed. Each agent was implemented as a modular function responsible for a specific task. The workflow followed a linear sequence, prompt-chained, where the output of each agent was passed as input to the next. 

The core components of the implementation included:

\begin{itemize}
    \item \textbf{Environment:} Python 3.11, with communication to the LLM handled through API calls to Ollama.
    \item \textbf{Prompt Templates:} Prompts were stored as templates and dynamically populated with relevant data.
    \item \textbf{Execution Logic:} A controller script coordinated the agents, invoking them sequentially and handling intermediate data transformation.
    \item \textbf{Logging and Validation:} Validation feedback was recorded for each execution to support iterative refinement.
\end{itemize}

The workflow follows a defined sequence:

\begin{enumerate}
    \item \textbf{Data Interpretation:} The \textit{Financial Data Analyst} (Agent 1) processes structured input data and converts it into natural language summaries, while also categorizing observed changes. This step serves to translate tabular financial data into a format that is more accessible for subsequent language-based reasoning, thereby improving the input quality for the downstream agents.

    \item \textbf{Trend Analysis:} The \textit{Business Analyst} (Agent 2) identifies significant trends, verifies consistency across product lines, and determines key influencing factors. By receiving pre-processed textual summaries from Agent 1, this agent is better equipped to perform analytical tasks that benefit from semantically processed input rather than raw data.

    \item \textbf{Report Generation:} The \textit{Report Writer} (Agent 3) compiles the analysis into a structured executive summary. The motivation behind this role is to ensure that the final output is clearly phrased, professionally formatted, and aligned with expectations for reporting style and tone, thus making it suitable for direct use or minimal revision in financial reports.

    \item \textbf{Validation:} The \textit{Financial Validator} (Agent 4) assesses whether the final summary is consistent with the input data and flags for factual inaccuracies. The rationale for including this role is to introduce a verification step that enhances the reliability of the output. While the current setup only performs validation passively, future iterations could enable feedback loops, where errors detected by the validator trigger corrections upstream. However, further development in that direction was constrained by computational limitations.
\end{enumerate}

\begin{figure}[H]
    \centering
    \includegraphics[width=1\linewidth]{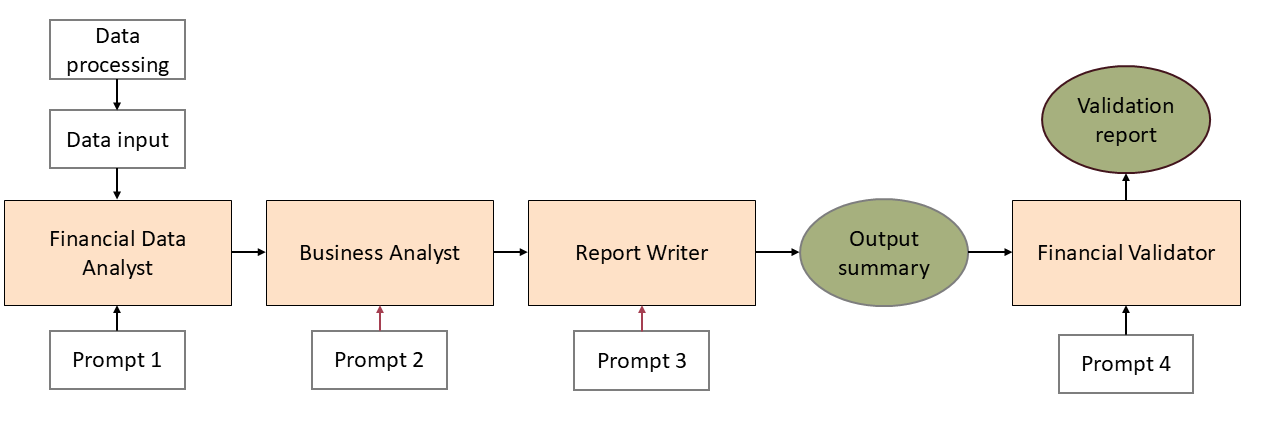}
    \caption{Figure illustrates the agent structure of the locally hosted LLM. Each box represents an agent with its own separate task, which it then sends over to the next agent.}
    \label{workflow1}
\end{figure}

\paragraph{Prompts}
Each agent was guided using a structured prompt composed of five parts: role, context, task, rules, and an expected output format. This format adheres to best practices in prompt design by providing the model with clear instructions and relevant contextual information, including the task's purpose, what it should and should not do, and what constitutes a successful output. The prompt for Agent one is presented below, and the rest can be found in Appendix \ref{tab:step1}.

\begin{table}[h]
\begin{tcolorbox}[colback=gray!10, colframe=gray!50, title=Step 1: Row-Level Summarization]

\textbf{Role:} You are a financial data analyst summarizing trends for the order intake trends for different product lines and regions

\hrulefill

\textbf{Context:}  
The dataset contains order intake changes across various product lines and regions between two periods. Each row represents the change, including its magnitude and the relative impact on the overall change.  

\hrulefill

\textbf{Table:} \texttt{\{table\_summary\}}

\hrulefill

\textbf{Task:}  
\begin{itemize}
    \item Process each row individually and convert the structured data into a natural language summary.  
    \item Clearly indicate whether the change is considered a major or minor impact.  
    \item Structure the output concisely while ensuring that key details are retained.  
\end{itemize}

\hrulefill

\textbf{Rules:}  
\begin{itemize}
    \item Use concise, structured sentences.  
    \item Do \textbf{not} include raw numbers, just directional trends.  
    \item Do \textbf{not} generate code.  
\end{itemize}

\hrulefill

\textbf{Example Output Format:}  
\texttt{[Product Line] in [Region] experienced a [major/minor] [increase/decrease].} 
\end{tcolorbox}

\captionsetup{type=table} 
\caption{Step 1: Row-Level Summarization}
\label{tab:step1.1}
\end{table}

\paragraph{Observations from Cycle 1}
The experimentation revealed several technical and usability challenges.

The initial testing was conducted using smaller models with 3 billion parameters. These models could run on a standard computer without significant resource demands. However, they had to be ruled out for continued experimentation, as they consistently struggled with following instructions and exhibited a high degree of hallucination.

For instance, when using LLaMA 3.2 and prompting the model to summarize trends from a table of financial data, it frequently responded with a Python script intended to calculate summary statistics, rather than providing a natural language summary. This behavior persisted even when explicitly instructed not to produce code and to instead give the response in natural language. The model often defaulted to describing how a script could be written, rather than directly providing the requested insights.

When using a chain of three language models, hallucination and not adhering to instructions became sometimes even more apparent. If one model in the chain deviates from the given instruction or hallucinates, that error is propagated and often amplified by downstream agents similar to the dynamics of a "whisper game."  For instance, when an upstream agent is instructed solely to summarize observed trends in the data, it may also infer potential causes or suggest follow-up actions based on the trends. These unintended inferences are then treated as factual input by downstream agents, which incorporate and build upon them in their respective responses. Figure \ref{fig:hallucination} illustrates a hallucinated output generated during such a chained interaction:

\begin{figure}[ht]
\centering
\fbox{%
\parbox{0.99\linewidth}{\ttfamily\small

\textbf{Summary} \\
Each product line has distinct features that cater to the unique needs of its respective region: \\
1. \textbf{CC Other} (Central and Eastern Europe): \\
\hspace*{1em}- Features: Customizable UI/UX design. \\
\hspace*{1em}- Benefits: Improved usability through tailored interface elements. \\[0.5em]
2. \textbf{CC Advanced Monitoring Disposables} (Central and Southern Hemisphere Development): \\
\hspace*{1em}- Features: Advanced analytics tools. \\
\hspace*{1em}- Benefits: Enhanced data analysis for better decision-making. \\[0.5em]
These features ensure that users in each region get a more personalized, effective product experience. This approach helps maximize user satisfaction and efficiency within their designated regions.
}}
\caption{Example of hallucinated LLM output}
\label{fig:hallucination}
\end{figure}

In this case, the output demonstrates a hallucinated response that resulted from a breakdown in the chained interaction. While the original task was to produce a concise summary of order intake trends, one of the upstream agents appears to have gone off track, generating fabricated content about product features and regional benefits that were neither part of the input nor requested. This misinformation was then passed along, with downstream agents treating it as valid input and building upon it. This resulted in an output that was entirely fabricated and disconnected from the input data.

Hallucination-related issues persisted when using larger models (8 billion parameters), although they appeared less frequently during experimentation. While the generated outputs often aligned with the input data and the expert created summaries, one example can be seen in table \ref{tab:summary_comparison}, one consistent challenges was the overall stability of the system. In several cases, the system failed to identify some of the more obvious trends and deviated much from the expected format.

\begin{table}[ht]
\centering
\begin{tabular}{|p{0.45\textwidth}|p{0.45\textwidth}|}
\hline
\textbf{Model-Generated Summary (Local LLM)} & \textbf{Expert-Written Summary} \\
\hline
Ventilation was a main growth driver, with major increases in US and APAC. Service saw an upward trend across all regions, with minor increases in each region. Anesthesia also experienced an upward trend, mainly driven by a major increase in EMEA. In contrast, Other and Advanced Monitoring Disposables saw downward trends, with minor decreases in all regions. 
& 
Ventilation in US as main growth driver. Service up in all regions. Anesthesia increasing, mainly in EMEA. Decrease from Other 3rd party products in Canada. Monitoring Disposables decreasing in China. \\
\hline
\end{tabular}
\caption{Example of model-generated summary compared to the corresponding one written by an expert.}
\label{tab:summary_comparison}
\end{table}

A challenge encountered was designing prompts that consistently guided the models to perform the intended tasks. Achieving the desired behavior, particularly when asking the system to identify relevant insights across multiple dimensions, proved complex. Similarly, crafting prompts that generalize well across variations in data distribution presented additional difficulties. 

The main constraint for the experimentation with running models locally was the computing demand required for inference when the necessary computational power was not available. Although the models used in the experiments were relatively small compared to many other LLMs, they still imposed hardware requirements that exceeded the computational capabilities typically available on standard laptops. As a result, further experimentation with this setup was impeded, and alternative solutions were subsequently investigated.

\subsection{Cycle 2: Cloud-based setup with GPT-4o}
To experiment with a commercial solution developed using OpenAI's GPT-4o model. This implementation aimed to explore the opportunities and challenges of using a large-scale model for generating natural language summaries of financial data.

The environment for this implementation was based on Python 3.11, utilizing OpenAI’s official Python SDK (version 1.0 or later) to facilitate interaction with the model. For the prompt design, we followed the same principles as in Cycle 1,  to define the model’s role and structure the task. These prompts were dynamically populated with context-specific data to ensure relevance and clarity. The same type of data pre-processing steps as used in Cycle 1 were employed in this iteration. To enhance readability and make the output suitable for direct inclusion in financial reports, the model’s responses were post-processed to split the summary into individual sentences and compiled into a CSV sheet.

The GPT-4o-based workflow is structured as follows:
\begin{enumerate} 
\item \textbf{System Role Definition:} The model is explicitly instructed to adopt the role of a financial analyst. This role prompts domain-specific behavior and constrains the model to maintain a professional and analytical tone aligned with financial reporting practices. The prompt emphasizes clarity, brevity, and factual accuracy, instructing the model to avoid speculation or numerical outputs.

\item \textbf{User Instruction and Input:} The user prompt follows a structured format inspired by instruction tuning techniques. It includes detailed task guidelines, style requirements, and formatting constraints, such as a maximum of four sentences and no numbers. It also draws inspiration from few-shot prompting by providing curated example outputs to guide the model toward the desired style and structure. The data block is injected into the prompt in a clearly marked input section. During experimentation, prompts without example outputs were explored. 

\item \textbf{Summary Generation:} The complete prompt is submitted to the GPT-4o API, which generates a concise, four-sentence qualitative summary. The model is guided to prioritize trends with the largest impact, identify consistent regional behaviors, and distinguish between positive and negative drivers while staying within strict output constraints.

\item \textbf{Post-processing and Output Export:} To improve readability, the generated summary is parsed so that each sentence is displayed on a separate line. The output is then saved to both a text file and a CSV file for further use in reporting or evaluation. \end{enumerate}

\begin{table}[H]
\begin{tcolorbox}[colback=gray!10, colframe=gray!50, title=Prompt to GPT-4o, after={\vspace{-1mm}}]
\textbf{Context:}\\
You are a financial analyst writing a financial report. Your task is to analyze and summarize changes in order intake data.

\hrulefill

\textbf{Instructions:}\\
You must write a concise summary in at most four sentences to complement the numerical data in a financial report.\\
Focus on clear, direct language and only state what is evident in the data—no speculation or recommendations.\\
Do not include numbers. Focus on trends, main drivers, and regions as specified.

\medskip
The data shows net sales changes segmented by product line and region, with relative contributions to the overall change in a business area.

\begin{itemize}
  \item Identify whether the overall trend is an increase or decrease.
  \item Detect the main positive and main negative drivers.
  \item Detect if there are product lines where the direction of change is consistent across all regions.
  \item Write a concise final qualitative summary about the detected trends.
  \item Start by mentioning the trend that has the biggest impact overall.
\end{itemize}

\textbf{Rules for the final summary:}
\begin{itemize}
  \item Maximum of four sentences.
  \item Do not speculate, do not include numbers.
  \item Do not include the overall trend in the summary since that is already understood.
  \item Use concise, factual phrasing as in the examples below.
\end{itemize}

\textbf{Expected output examples:}
\begin{itemize}
  \item \texttt{All product lines up. [Product Line] in [Region] as main growth driver. [Product Line] increasing in [Region]. Increase from [Product Line] in [Region]. [Product Line] up in all regions.}
  \item \texttt{[Product Line] and [Product Line] decreasing in [Region]. [Product Line] in [Region] as main detractor, partly offset by [Product Line] in [Region].}
  \item \texttt{[Product Line] in [Region] as main detractor. Increase from [Product Line] in [Region].}
  \item \texttt{[Product Line] up in all regions. [Product Line] down in [Region]. [Product Line] up in [Region].}
\end{itemize}

\textbf{Input:} \\
\texttt{Data: \{data\_block\}}
\end{tcolorbox}

\captionsetup{type=table}
\caption{Prompt for Generating Order Intake Summary}
\label{GPT4oPrompt}
\end{table}
\newpage

\paragraph{Observations from Cycle 2}
Setting up this requires minimal setup and configuration during experimentation. In our tested use cases, summarizing financial trends, the model appears to capture most underlying trends without requiring the same level of task decomposition. An example of the generated and expert-written summary can be seen in Table \ref{tab:summary_cloud}. However, it sometimes misses important insights. 

\begin{table}[H]
\centering
\begin{tabular}{|p{0.45\textwidth}|p{0.45\textwidth}|}
\hline
\textbf{Model-Generated Summary} & \textbf{Expert-Written Summary} \\
\hline
Beta Bags and Consumables in EMEA as main growth driver. Major increases in Service in the US and Fluid Pathway in the US. Sterilization in EMEA as main detractor, with significant decreases also in China and the Americas. Bio Reactors up in EMEA and the US, but down in China and APAC. &
BetaBags increasing in all regions, primarily in EMEA. Service up in all regions. Increase from Bio Reactors in US and EMEA. Fluid Pathway up in all regions, mainly in US. Decline from Sterilization in all regions. Isolation down in all regions, primarily EMEA. \\
\hline
\end{tabular}
\caption{Example summary from GPT-4o compared with the expert written one.}
\label{tab:summary_cloud}
\end{table}

When provided with example summaries to guide format and tone, GPT-4o often mirrored these examples too rigidly, resulting in outputs that followed the structure but also occasionally ignored important variations in the input data. In contrast, when no format example was provided, the model captured the underlying trends more accurately but produced summaries that did not adhere to the desired format. To get the model to produce the desired output, clear and representative examples were necessary.

Deployment of a high-capacity model such as GPT-4o would be a challenge due to its significant computational requirements. On-premise deployment was not feasible within the constraints of the available hardware, necessitating the use of third-party hosting solutions. This setup introduced considerations related to data privacy, regulatory compliance, and dependence on external providers, which emerged as challenges when integrating LLMs in real-world financial workflows.

\subsection{Evaluation}
\label{Evaluation of summarization tasks}
The evaluation compared the quality of two AI-based summarization approaches, one using GPT-4o and the other a local LLM solution to assess their effectiveness in supporting financial reporting tasks. Each model generated 20 summaries based on the same underlying financial data, which were then compared to corresponding human-written summaries used in practice. Feedback was gathered from two professional users with experience in financial reporting, who evaluated each model’s output across five key dimensions: usefulness, accuracy, clarity, trust, and integration into existing reporting workflows.

\paragraph{Clarity and Language}

According to the participants, both the GPT-4o and the local LLM models produced outputs that were professionally formulated and suitable for use in financial contexts. Participant 1 and Participant 2 both described GPT-4o’s language as concise and aligned with the tone typically expected in financial reporting, particularly for presentation purposes. The phrasing was characterized as consistent with the style used by financial domain experts.

The local LLM was also described as maintaining a professional tone but generating more descriptive outputs. Participant 1 stated that “the local LLM offers a more descriptive summary that is easier to understand, but it occasionally includes unnecessary details.” Based on this feedback, the local model’s output was considered potentially suitable for internal use due to its format and level of detail.

Both participants observed that, in some cases, the presentation of information reduced clarity. GPT-4o was reported to occasionally include both ``main drivers'' and ``main detractors'' within the same summary section. In situations where a single, dominant factor was expected, this structure was perceived as potentially ambiguous. According to the participants, this occasionally made it difficult to determine whether the overall trend was positive or negative. They also noted that positive and negative developments were sometimes presented without a clearly defined structure. The following excerpt was cited by one participant as an example of this issue:

\begin{quote}
\textit{“Disposables ECLS in the US and EMEA as main growth drivers. Hardware increasing in all regions. Disposables Surgical Perfusion in EMEA as main detractor. Service up in the US and EMEA.”}
\end{quote}

Additionally, both models were noted to have a tendency to repeat themselves, mentioning the same product line multiple times within the same product area. Participant 1 commented: \textit{"Both the local LLM and GPT-4o model sometimes include the same product line multiple times in a product area that I would not comment on myself, as the focus should be on the most significant changes."}

\paragraph{Accuracy and Reliability}
Both participants identified aspects related to the accuracy and reliability of the summaries. Feedback indicated that the summaries sometimes omitted key information or included unnecessary details. Participant 1 summarized this observation as follows: “The summaries are helpful as a foundation, but they do not always explain the trends accurately. GPT-4o sometimes omits key drivers or mixes positive and negative changes in a way that makes interpretation difficult. Additionally, it had a tendency to misprioritize the most impactful changes.”

Feedback related to the local LLM model suggested that while it captured broader trends, its descriptive style occasionally led to the inclusion of less relevant details. This was noted as potentially limiting the clarity of the key drivers, though it was not necessarily characterized as inaccurate.

Regarding overall trust and reliability, both participants stated that manual review would still be necessary before using the summaries in formal reporting. According to participant 2, “The summaries still need manual input for the comments of the reporting,” and the outputs were described as more appropriate for internal use.

Both participants also noted a lack of transparency in how content was selected and structured by the models. Participant 2 remarked that the rationale behind which elements were emphasized was not always clear, making it difficult to assess the reasoning behind specific summary components. Additionally, both experts highlighted the lack of numerical values in the generated outputs as a limitation, stating that the inclusion of such figures would help contextualize changes and improve evaluative precision.

Participant 1 also reported instances of unclear or incorrect regional attributions, noting that such errors could result in misinterpretation when used in a financial reporting context.

While the feedback indicated certain limitations in accuracy and reliability, both participants noted that the model-generated summaries could serve as a helpful starting point for drafting financial commentary, particularly if supplemented with manual adjustments.

\paragraph{Summary of the Evaluation Findings}
The evaluation indicated challenges across five themes. GPT-4o was preferred for reporting purposes due to its concise and professional language, while the Local LLM solution shows some potential to be suitable for internal analysis. Summaries from both models would require manual refinement, often including redundant or irrelevant content. Clarity was generally intense, particularly in GPT-4o, but both models occasionally mixed trends or listed multiple ``main drivers.'' Accuracy issues were observed, including the omission of key changes and incorrect regional attributions. Structural inconsistencies and the absence of numerical values affected prioritization. While both models facilitated faster draft creation, their use in formal reporting remained limited due to concerns about trust and accuracy.

\begin{table}[H]
\centering
\label{tab:challenges_summary}
\begin{tabular}{|p{4cm}|p{10cm}|}
\hline
\textbf{Theme} & \textbf{Identified Challenges} \\ \hline
\textbf{Usability} & Summaries require manual refinement due to redundancy and inclusion of less relevant content. \\ \hline
\textbf{Language} & Sounds professional, but occasional inconsistencies in phrasing and emphasis affect interpret ability. \\ \hline
\textbf{Accuracy} & Some omissions of important information and inconsistencies in the trends. \\ \hline
\textbf{Trust and Reliability} & Limited confidence in summaries due to occasional missed trends. \\ \hline
\textbf{Workflow Integration} & Models support early-stage drafting but are not yet dependable for finalized reporting without human validation. \\ \hline
\end{tabular}
\caption{Summary of Key Challenges by Evaluation Theme}
\end{table}

\subsection{Refinement of Cycle 2: Prompt-Chained Approach}
Following the evaluation of the solution developed in Cycle 2, several limitations were identified, and it was concluded that refinements would be necessary if the solution were to produce output suitable for a final report. These included the model's occasional failure to adhere to formatting constraints, difficulties in handling multiple concurrent instructions, and challenges in logical consistency within the summary. 

To address these issues, a prompt-chaining strategy was introduced as a refinement of the Cycle 2 solution. Rather than using a single prompt to both interpret and summarize the financial data, the task was split into two distinct steps.

\begin{itemize}
    \item \textbf{Step 1 – Analyst Prompt:} The model first acted as a financial data analyst, interpreting the data and outputting a structured breakdown of product-line level changes. The analyst’s output included details on key regions, consistency across regions, and each product line’s impact on the overall trend.
    
    \item \textbf{Step 2 – Report Writer Prompt:} The analysis was then passed as input to a second prompt, where the model adopted the role of a report writer. This step focused purely on writing the summary in the specified format. Based on the feedback from the evaluation, some instructions were added and made stricter to make outputs clearer and closer to the desired format used in reporting. 
    For example, a product line in a single summary or a summary can not include both a driver that is a main detractor and a main growth driver at the same time. 
\end{itemize}

\begin{figure}[h]
\centering
\includegraphics[width=1.2\linewidth]{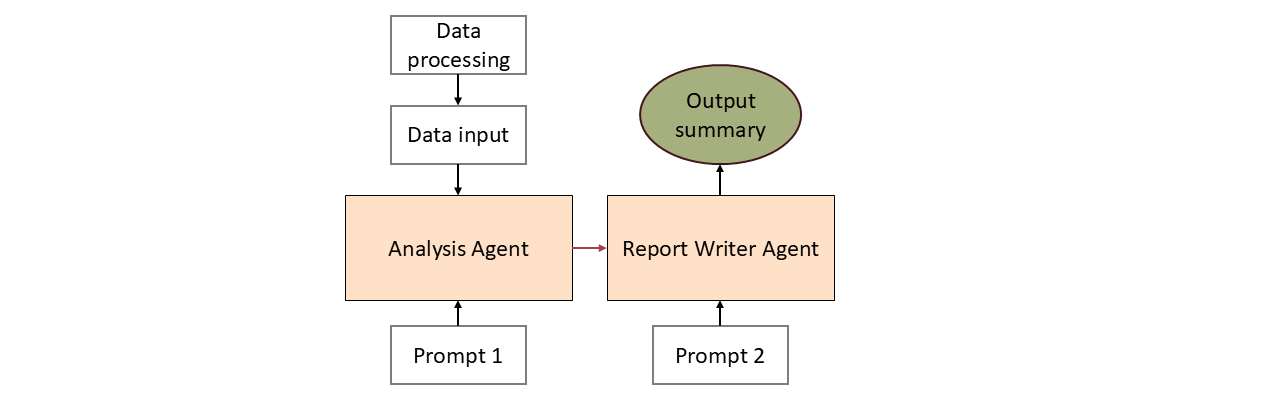}
\caption{Two-agent system with analysis and summary generation.}
\label{PromptchainedAproach}
\end{figure}

This separation of concerns was intended to reduce the cognitive load placed on the model in a single prompt and to improve adherence to formatting rules and logical constraints. 

\paragraph{Observations}
The prompt-chained approach led to improvements in how the model interpreted financial data and structured its analysis. The analytical output produced in the first stage generally corresponded well with the underlying financial trends, indicating that the model was able to extract and organize relevant information at the product-line and regional level with a high degree of alignment to the data.

However, several limitations still persisted in the second stage, where the model generated the final report-ready summaries. Although improved, one limitation was the continued difficulty in prompting the model to consistently follow highly specific instructions, particularly those resembling conditional logic (e.g., “if X, then Y”). This issue became especially apparent when multiple constraints needed to be satisfied simultaneously. For example, summaries occasionally mentioned a main growth driver even when the overall direction of change was negative, despite the prompt explicitly instructing the model to avoid such statements under these conditions. Similarly, some summaries included repeated references to the same product line, contrary to the instruction that each product line must be mentioned only once.

In addition, the commentary generated in the second stage did not always align with the preceding structured analysis. Although the analysis step correctly identified key product-line and regional trends, these elements were sometimes omitted, rephrased inaccurately, or contradicted in the final summary. 

These findings suggest that while dividing the summarization task into sequential roles reduces some of the cognitive load placed on the model, the challenge of translating structured intermediate output into precise, constraint-adherent final summaries remains. This highlights the difficulty for LLMs in reliably applying logic and formatting rules in text generation tasks within financial reporting, when multiple nuanced instructions must be satisfied concurrently.

\section{Discussion}
This section presents the key findings from our iterative development and evaluation of open-source and commercial LLM-based solutions in the context of financial report analysis workflows. The discussion is structured around the two research questions: first examining the primary software engineering challenges encountered when integrating open and commercial LLMs (RQ1), followed by a comparative analysis of different implementation approaches and their implications for system design and deployment in financial reporting (RQ2).

\subsection{Challenges When Designing Different LLM Solutions}
This section addresses RQ1 by identifying and discussing the primary software engineering challenges encountered when designing and integrating LLMs into financial report analysis workflows. Based on our iterative development and evaluation process, three key challenges emerged: 1. task-specificity in prompt and system design, 2. variability and complexity in data preparation, and 3. constraints arising from data sensitivity.

\paragraph{Task Specificity in Financial LLM Applications}
One challenge emerging from the study is the high degree of specificity in financial reporting use cases and the impact it has on the design of LLM-based systems. The summaries intended for inclusion in reports must follow particular formats, be concise, precise, and unambiguous to align with domain expectations and support accurate interpretation. 

Even with clear and explicitly worded prompts, the solutions struggle to follow set rules for output. Getting an LLM to produce a summary in a specific format posed a persistent challenge, particularly where certain phrasing should be avoided if specific conditions are met. While the language used generally aligns with financial reporting tone, the summaries frequently became ambiguous to readers with domain knowledge. 

The introduction of a prompt-chaining strategy improved the overall structure and alignment for the analytical part of the task. However, errors still appeared in the final summarization stage, where output sometimes failed to reflect key points identified in the structured analysis. This suggests that LLMs struggle to handle many specific instructions simultaneously. While structured prompting strategies show improvements in model reasoning for financial prediction tasks \cite{Kim2024}, these techniques do not fully apply to summarization tasks that impose strict constraints on phrasing, tone, and structure.

This reflects a broader system-level challenge identified in previous research. Mahr et al. \cite{mahr2024reference} frame prompt engineering as only one component within a larger architectural approach, emphasizing the importance of complementary mechanisms such as evaluation loops and verification steps to ensure reliable behavior in precision-critical applications. Our findings support this view, particularly in end-to-end automated financial reporting workflows, where strict adherence to formatting and logical rules is required.

\paragraph{Data Preparation and Variability}
Another obstacle lies in the complexity of preparing and processing data used during inference. Financial reporting often requires accessing data from multiple systems in varied formats --- spreadsheets, presentation decks, ERP systems, databases, or verbal communications from meetings. This diversity presents an obstacle to developing generalizable and robust LLM workflows.

In this study, input data were preprocessed into a standardized format suitable for model inference through a custom script. While effective for the specific scenario, the solution was tightly coupled to a fixed input structure and lacked generalization across other similar tasks.

One potential approach to increasing system flexibility is using tool-enabled agents. LLMs can be integrated with external tools for data extraction and transformation, dynamically generating and executing scripts to retrieve relevant information. Frameworks such as ReAct show how chain‑of‑thought reasoning can be interleaved with tool calls \cite{yao2023react}, while orchestration patterns demonstrate how tasks can be decomposed and delegated to specialized worker agents \cite{anthropic2024agents}. Advanced models like OpenAI's o3 series embed function calling and code execution capabilities directly, further reducing dependence on static prompts \cite{openai2025o3}.

Multimodal capabilities may also help address format challenges. Models like GPT-4o can interpret textual, verbal, and visual information in a single context window \cite{jaech2024openai}, potentially handling varied inputs like scanned documents or audio transcripts without explicit manual conversion.

\paragraph{Constraints from Data Sensitivity}
Financial reports often contain confidential and proprietary information, imposing specific requirements on data handling and model deployment. While such concerns can be addressed through vendor agreements or private cloud configurations, these arrangements limit flexibility in choosing models or services, potentially preventing adoption of more suitable alternatives. This theme appears in literature where Tavasoli et al. \cite{tavasoli2025responsible} discuss trade-offs between vendor-managed and internal deployments, highlighting tensions between data control, compliance, and operational efficiency, as well as vendor lock-in risks.

\subsection{Analysis of Implementation Approaches}
Following the discussion of main engineering challenges regarding RQ1, attention now turns to RQ2, which concerns comparing different implementation approaches. Two solutions were developed and evaluated: a locally deployed one and one using GPT-4o. Expert feedback assessed output quality, reliability, and feasibility for practical use.

\paragraph{Implementation with Local LLM Models}
The main motivation for exploring local, open-source LLMs was their potential to meet key requirements within the financial domain: full control over data, privacy assurance, and reduced dependency on external services. During experimentation, several limitations with small-scale models became apparent. These models frequently struggled with instruction following, hallucinated content, and occasionally returned code instead of natural language.

To address these limitations, a multi-agent setup was implemented, improving output coherence and reducing hallucinations while providing a more modular interaction flow. Yet it introduced new challenges, as early errors in the pipeline were sometimes propagated downstream, compounding deviations. These challenges reflect broader limitations in designing multi-agent LLM systems, often rooted in systemic design flaws rather than model capacity alone \cite{pan2025multiagent}.

While the system could produce summaries that often aligned with underlying data, two key issues persisted: difficulty in prompting consistent use of expected tone and phrasing for formal financial reporting, and limited overall reliability as some summaries failed to capture key trends from input data.

\paragraph{Implementation with GPT-4o}
GPT-4o demonstrated ability to identify trends without extensive task decomposition, suggesting that powerful general-purpose LLMs can capture domain-relevant patterns with relatively abstract task definitions. Based on expert feedback, GPT-4o retained business-oriented linguistic patterns and generated summaries similar to those written by professionals. However, the model suffered from structural issues, occasionally including multiple "main drivers" in a single summary or mixing negative and positive trends.

When example summaries were provided to guide structure and tone, the model tended to imitate examples too rigidly, sometimes reproducing phrasing from prompts even when it conflicted with actual input data. A prompt chaining strategy led to improvements in data extraction and interpretation, but the final summaries were often misaligned with previous analysis, especially when prompts included conditional logic.

\paragraph{Comparisons of the Approaches}
Expert feedback revealed important differences between approaches in terms of output quality, clarity, and structure. The locally deployed LLM produced verbose, descriptive outputs offering potential value for internal documentation but lacking conciseness required for formal reporting. This variability significantly affects software complexity, as filtering content, enforcing structure, and correcting phrasing require additional post-processing logic, increasing codebase intricacy and reducing maintainability.

GPT-4o produced summaries that more consistently matched expected tone and structure, reducing development effort and simplifying system maintenance. However, occasional logical inconsistencies still require engineering interventions such as verification modules. A shared challenge across both approaches is lack of interpretability, making it difficult to debug inconsistencies or validate results, complicating error handling, testing, and quality assurance.

The findings suggest that while both models can produce draft-quality summaries, they fall short of delivering report-ready material suitable for external financial reports. The local model demands greater engineering investment to control behavior, whereas GPT-4o offers a more usable foundation with residual logic and reliability issues. These trade-offs underscore the need to evaluate LLM solutions based not only on output quality but also on their implications for code maintainability, complexity, and reliability in financial software systems.

\subsection{Implications}
Although LLM capabilities have been extensively explored in recent research, the software engineering perspective on how to build, integrate, and maintain open-source or commercial LLM systems remains underdeveloped. Existing literature offers limited guidance on engineering practices and tool support for developing reliable, maintainable, and auditable LLM-based applications in complex, domain-specific environments.

This study has investigated the challenges of integrating LLMs into financial reporting workflows, as well as two integration approaches. Through this, several key software engineering challenges have emerged, including engineering for task-specific tasks, coordination within multi-agent systems, and trade-offs between deployment options. These findings highlight the need for more structured methods, evaluation practices, and architectural support for implementing LLMs in real-world contexts like finance.
Although approaches like Chain of Thought (CoT) prompting have shown effectiveness in financial reasoning tasks, improving prediction accuracy \cite{Kim2024}, and organizations like Anthropic \cite{anthropic2024promptguide} have outlined best practices for prompt design, the findings suggest that these techniques may be insufficient for tasks with highly specific output requirements. In structured financial reporting, prompts must not only guide the model toward the correct content but also enforce precise phrasing, tone, and formatting, often under conditional rules. This level of specificity is a challenge.

The exploration of multi-agent setups showed that architectural modularity improves task decomposition and clarifies task boundaries, but also introduces fragility when early-stage components produce faulty outputs. Errors tend to propagate across chained agents, particularly when verification steps are weak or absent. These priorities align with the architectural taxonomy proposed by Lu et al. \cite{lu2024taxonomy}, which identifies prompt chaining, verifier modules, and coordination layers as critical for ensuring robustness in LLM workflows.

The comparison between local and commercial cloud-based models highlighted trade-offs between control, cost, and ease of integration. Local models provide full control over data and deployment environments, which is crucial for meeting strict privacy or compliance requirements. However, they demand significant engineering effort and infrastructure. In contrast, cloud-based models like GPT-4o offer ease of use and strong out-of-the-box performance but raise concerns about data sensitivity, compliance risks, and vendor lock-in.

\subsection{Limitations and Threats to Validity}
A limitation affecting internal validity is the limited number of domain experts involved in the evaluation. This small sample size was primarily due to limited access to individuals with the necessary expertise. As a result, only qualitative expert assessments were feasible, which may introduce subjectivity and reduce the consistency of evaluation outcomes. While the insights provided were valuable for assessing the solutions relevance and usability, the limited pool of evaluators constrains the robustness of the conclusions drawn.

Regarding construct validity, the evaluation relies on qualitative expert assessments rather than quantitative metrics, due to the structured and domain-specific nature of the generated outputs. Common metrics such as BLEU or ROUGE are not well-suited to capture correctness, regulatory compliance, or business relevance in financial reporting. By using expert judgment aligned with domain expectations, the research aimed to more accurately assess the artifacts real-world utility. However, this approach introduces a degree of subjectivity and limits comparability with other systems.

A practical limitation affecting external validity is the use of transformed and old financial data, introduced due to confidentiality constraints. Although the data was modeled to reflect real-world characteristics, actual organizational variability and data noise may influence performance in practice. Furthermore, the artifact was designed and evaluated within the context of a specific financial reporting workflow at a single company. This narrow context limits the extent to which the findings can be generalized to other industries, organizational structures, or use cases.

Finally, reliability may be impacted by factors such as model non-determinism and infrastructure differences, as the system involves prompt engineering and multi-agent orchestration of LLMs.

\section{Conclusions and Future Work}
This study examined the integration of open-source and commercial LLMs into domain-specific report analysis workflows, with a primary focus on the software engineering challenges that arise when designing these solutions. Through a Design Science approach, both local and cloud-based LLM systems were designed and evaluated for automating summarization tasks in a financial context. The study reveals that while LLMs hold potential in automating reporting tasks, they present significant engineering challenges including managing high task specificity, enforcing strict formatting rules, processing heterogeneous and sensitive data inputs, and ensuring traceability and explainability of outputs. These challenges are especially pronounced in precision-critical domains like financial reporting, where inconsistencies or ambiguity can compromise the usefulness and reliability of the output.

While full end-to-end automation of formal financial reporting remains challenging, promising progress has been demonstrated in using LLMs to support and partially automate such workflows.

Future work could explore modular and rule-aware prompt templates and methods for integrating conditional logic to address task specificity challenges, while investigating tooling for prompt versioning, documentation, and performance tracking as Weber et al. \cite{weber2024llm} suggest, i.e., treating prompts as software assets. Additionally, research should focus on formalizing coordination protocols and enhancing fault isolation in multi-agent setups, building on Lu et al.'s \cite{lu2024taxonomy} architectural insights, alongside exploring hybrid deployment strategies that combine local processing for sensitive tasks with external models for complex operations, as practical implementation approaches for such frameworks remain underexplored.

\section*{Declaration of generative AI and AI-assisted technologies in the writing process}
During the preparation of this work the author(s) used ChatGPT and Claude-ai in order to clarify and improve the language. After using this tool/service, the author(s) reviewed and edited the content as needed and take(s) full responsibility for the content of the publication.

\bibliographystyle{elsarticle-num}
\bibliography{references}

\begin{thebibliography}{10}
\expandafter\ifx\csname url\endcsname\relax
  \def\url#1{\texttt{#1}}\fi
\expandafter\ifx\csname urlprefix\endcsname\relax\def\urlprefix{URL }\fi
\expandafter\ifx\csname href\endcsname\relax
  \def\href#1#2{#2} \def\path#1{#1}\fi

\bibitem{vaswani2017attention}
A.~Vaswani, N.~Shazeer, N.~Parmar, J.~Uszkoreit, L.~Jones, A.~N. Gomez,
  L.~Kaiser, I.~Polosukhin, Attention is all you need, in: Advances in Neural
  Information Processing Systems, Vol.~30, 2017, pp. 5998--6008.

\bibitem{MENG2024109713}
X.~Meng, X.~Yan, K.~Zhang, D.~Liu, X.~Cui, Y.~Yang, M.~Zhang, C.~Cao, J.~Wang,
  X.~Wang, J.~Gao, Y.-G.-S. Wang, J.-m. Ji, Z.~Qiu, M.~Li, C.~Qian, T.~Guo,
  S.~Ma, Z.~Wang, Z.~Guo, Y.~Lei, C.~Shao, W.~Wang, H.~Fan, Y.-D. Tang,
  \href{https://doi.org/10.1016/j.isci.2024.109713}{The application of large
  language models in medicine: A scoping review}, iScience 27~(5) (2024)
  109713.
\newblock \href {http://dx.doi.org/10.1016/j.isci.2024.109713}
  {\path{doi:10.1016/j.isci.2024.109713}}.
\newline\urlprefix\url{https://doi.org/10.1016/j.isci.2024.109713}

\bibitem{Baierl2023}
J.~D. Baierl, Applications of large language models in education: Literature
  review and case study, Master's thesis, University of California, Los
  Angeles, available from: https://escholarship.org/uc/item/6kf0r28s (2023).

\bibitem{bespokeLegal}
T.~Jayakumar, F.~Farooqui, L.~Farooqui, Large language models are legal but
  they are not: Making the case for a powerful legalllm, in: Proceedings of the
  Natural Legal Language Processing Workshop 2023, 2023, pp. 223--229.

\bibitem{liang2024encouragingdivergentthinkinglarge}
T.~Liang, Z.~He, W.~Jiao, X.~Wang, Y.~Wang, R.~Wang, Y.~Yang, S.~Shi, Z.~Tu,
  Encouraging divergent thinking in large language models through multi-agent
  debate, in: Proceedings of the 2024 Conference on Empirical Methods in
  Natural Language Processing, 2024, pp. 17889--17904.

\bibitem{alnaqbi2024enhancing}
H.~Al~Naqbi, Z.~Bahroun, V.~Ahmed,
  \href{https://www.mdpi.com/2071-1050/16/3/1166}{Enhancing work productivity
  through generative artificial intelligence: A comprehensive literature
  review}, Sustainability 16~(3).
\newblock \href {http://dx.doi.org/10.3390/su16031166}
  {\path{doi:10.3390/su16031166}}.
\newline\urlprefix\url{https://www.mdpi.com/2071-1050/16/3/1166}

\bibitem{Yu2024}
Y.~Yu, P.~Tiwari, \href{https://arxiv.org/abs/2410.21338}{Finteamexperts: Role
  specialized moes for financial analysis}, arXiv preprint (2024).
\newblock \href {http://arxiv.org/abs/2410.21338} {\path{arXiv:2410.21338}}.
\newline\urlprefix\url{https://arxiv.org/abs/2410.21338}

\bibitem{Akcan2025}
S.~Akcan~Yetgin, H.~Altas, Analyzing the corporate business intelligence
  impact: A case study in the financial sector, Applied Sciences 15 (2025)
  1012.
\newblock \href {http://dx.doi.org/10.3390/app15031012}
  {\path{doi:10.3390/app15031012}}.

\bibitem{Wu2023}
S.~Wu, O.~Irsoy, S.~Lu, V.~Dabravolski,
  \href{https://arxiv.org/abs/2303.17564}{Bloomberggpt: A large language model
  for finance}, arXiv preprint (2023).
\newblock \href {http://arxiv.org/abs/2303.17564} {\path{arXiv:2303.17564}},
  \href {http://dx.doi.org/10.48550/arXiv.2303.17564}
  {\path{doi:10.48550/arXiv.2303.17564}}.
\newline\urlprefix\url{https://arxiv.org/abs/2303.17564}

\bibitem{privacy_preserving}
E.~A. Abbe, A.~E. Khandani, A.~W. Lo,
  \href{http://www.jstor.org/stable/23245506}{Privacy-preserving methods for
  sharing financial risk exposures}, The American Economic Review 102~(3)
  (2012) 65--70.
\newline\urlprefix\url{http://www.jstor.org/stable/23245506}

\bibitem{araci2019finbertfinancialsentimentanalysis}
D.~Araci, \href{https://arxiv.org/abs/1908.10063}{Finbert: Financial sentiment
  analysis with pre-trained language models} (2019).
\newblock \href {http://arxiv.org/abs/1908.10063} {\path{arXiv:1908.10063}}.
\newline\urlprefix\url{https://arxiv.org/abs/1908.10063}

\bibitem{uchitel2024scoping}
S.~Uchitel, M.~Chechik, M.~Di~Penta, B.~Adams, N.~Aguirre, G.~Bavota,
  D.~Bianculli, K.~Blincoe, A.~Cavalcanti, Y.~Dittrich, F.~Ferrucci, Scoping
  software engineering for ai: The tse perspective, Institute of Electrical and
  Electronics Engineers.

\bibitem{jm3}
D.~Jurafsky, J.~H. Martin,
  \href{https://web.stanford.edu/~jurafsky/slp3/}{Speech and Language
  Processing: An Introduction to Natural Language Processing, Computational
  Linguistics, and Speech Recognition with Language Models}, 3rd Edition,
  Manuscript, Stanford University, 2025, online manuscript released January 12,
  2025.
\newline\urlprefix\url{https://web.stanford.edu/~jurafsky/slp3/}

\bibitem{Weizenbaum1966}
J.~Weizenbaum, {ELIZA}—a computer program for the study of natural language
  communication between man and machine, Communications of the ACM 9~(1) (1966)
  36--45.
\newblock \href {http://dx.doi.org/10.1145/365153.365168}
  {\path{doi:10.1145/365153.365168}}.

\bibitem{chen1999empirical}
S.~F. Chen, J.~Goodman, An empirical study of smoothing techniques for language
  modeling, Computer Speech \& Language 13~(4) (1999) 359--394.

\bibitem{hochreiter1997long}
S.~Hochreiter, J.~Schmidhuber, Long short-term memory, Neural computation 9~(8)
  (1997) 1735--1780.

\bibitem{devlin2018bert}
J.~Devlin, Bert: Pre-training of deep bidirectional transformers for language
  understanding, arXiv preprint arXiv:1810.04805.
\newblock \href {http://arxiv.org/abs/1810.04805} {\path{arXiv:1810.04805}}.

\bibitem{radford2018improving}
A.~Radford, K.~Narasimhan, T.~Salimans, I.~Sutskever, Improving language
  understanding by generative pre-training, Technical report, OpenAI (2018).

\bibitem{brown2020language}
T.~Brown, B.~Mann, N.~Ryder, M.~Subbiah, J.~D. Kaplan, P.~Dhariwal,
  A.~Neelakantan, P.~Shyam, G.~Sastry, A.~Askell, et~al., Language models are
  few-shot learners, Advances in neural information processing systems 33
  (2020) 1877--1901.

\bibitem{anthropic2024promptguide}
{Anthropic},
  \href{https://docs.anthropic.com/en/docs/build-with-claude/prompt-engineering/be-clear-and-direct}{Be
  clear, direct, and detailed [internet]}, san Francisco, CA: Anthropic; 2024
  [cited 2025 May 8]. Available from:
  https://docs.anthropic.com/en/docs/build-with-claude/prompt-engineering/be-clear-and-direct
  (2024).
\newline\urlprefix\url{https://docs.anthropic.com/en/docs/build-with-claude/prompt-engineering/be-clear-and-direct}

\bibitem{wei2022chain}
J.~Wei, X.~Wang, D.~Schuurmans, M.~Bosma, F.~Xia, E.~Chi, Q.~V. Le, D.~Zhou,
  et~al., Chain-of-thought prompting elicits reasoning in large language
  models, Advances in neural information processing systems 35 (2022)
  24824--24837.

\bibitem{anthropic2025claude37}
{Anthropic}, \href{https://www.anthropic.com/claude/sonnet}{Claude 3.7 sonnet
  [internet]}, san Francisco, CA: Anthropic; 2025 [cited 2025 May 7]. Available
  from: https://www.anthropic.com/claude/sonnet (2025).
\newline\urlprefix\url{https://www.anthropic.com/claude/sonnet}

\bibitem{google2025gemini25pro}
{Google Cloud},
  \href{https://cloud.google.com/vertex-ai/generative-ai/docs/models/gemini/2-5-pro}{Gemini
  2.5 pro [internet]}, mountain View, CA: Google Cloud; 2025 [cited 2025 May
  7]. Available from:
  https://cloud.google.com/vertex-ai/generative-ai/docs/models/gemini/2-5-pro
  (2025).
\newline\urlprefix\url{https://cloud.google.com/vertex-ai/generative-ai/docs/models/gemini/2-5-pro}

\bibitem{liu2024lost}
N.~F. Liu, K.~Lin, J.~Hewitt, A.~Paranjape, M.~Bevilacqua, F.~Petroni,
  P.~Liang, Lost in the middle: How language models use long contexts,
  Transactions of the Association for Computational Linguistics 12 (2024)
  157--173.

\bibitem{peng2023agentbench}
B.~Peng, C.-S. Wu, A.~Chronopoulos, X.~Tang, H.~Elsahar, J.-Y. Kao, L.~Liden,
  M.~Galley, J.~Gao, \href{https://arxiv.org/abs/2308.11458}{{AgentBench:
  Evaluating foundation models as agents}}, arXiv preprint arXiv:2308.11458.
\newline\urlprefix\url{https://arxiv.org/abs/2308.11458}

\bibitem{yao2023react}
S.~Yao, J.~Zhao, D.~Yu, N.~Du, I.~Shafran, K.~Narasimhan, Y.~Cao,
  \href{https://arxiv.org/abs/2210.03629}{React: Synergizing reasoning and
  acting in language models}, in: Proceedings of the 11th International
  Conference on Learning Representations (ICLR), 2023, arXiv preprint
  arXiv:2210.03629.
\newline\urlprefix\url{https://arxiv.org/abs/2210.03629}

\bibitem{anthropic2024agents}
{Anthropic},
  \href{https://www.anthropic.com/research/building-effective-agents}{Building
  effective agents [internet]}, san Francisco, CA: Anthropic; 2024 [cited 2025
  Feb 19]. Available from:
  https://www.anthropic.com/research/building-effective-agents (2024).
\newline\urlprefix\url{https://www.anthropic.com/research/building-effective-agents}

\bibitem{pan2025multiagent}
M.~Z. Pan, M.~Cemri, L.~A. Agrawal, S.~Yang, B.~Chopra, R.~Tiwari, K.~Keutzer,
  A.~Parameswaran, K.~Ramchandran, D.~Klein, et~al., Why do multiagent systems
  fail?, in: ICLR 2025 Workshop on Building Trust in Language Models and
  Applications, 2025.

\bibitem{jaech2024openai}
A.~Jaech, A.~Kalai, A.~Lerer, A.~Richardson, A.~El-Kishky, A.~Low, A.~Helyar,
  A.~Madry, A.~Beutel, A.~Carney, et~al., Openai o1 system card, arXiv preprint
  arXiv:2412.16720.

\bibitem{anthropic2025claude37card}
{Anthropic}, \href{https://www.anthropic.com/news/claude-3-7-sonnet}{Claude 3.7
  sonnet system card [internet]}, Tech. rep., Anthropic PBC, san Francisco, CA:
  Anthropic PBC; 2025 Feb [cited 2025 May 7]. Available from:
  https://www.anthropic.com/news/claude-3-7-sonnet (2025).
\newline\urlprefix\url{https://www.anthropic.com/news/claude-3-7-sonnet}

\bibitem{shahriar2024putting}
S.~Shahriar, B.~D. Lund, N.~R. Mannuru, M.~A. Arshad, K.~Hayawi, R.~V.~K.
  Bevara, A.~Mannuru, L.~Batool, Putting gpt-4o to the sword: A comprehensive
  evaluation of language, vision, speech, and multimodal proficiency, Applied
  Sciences 14~(17) (2024) 7782.

\bibitem{frantar2022gptq}
E.~Frantar, S.~Ashkboos, T.~Hoefler, D.~Alistarh, Gptq: Accurate post-training
  quantization for generative pre-trained transformers, arXiv preprint
  arXiv:2210.17323.

\bibitem{ma2023llm}
X.~Ma, G.~Fang, X.~Wang, Llm-pruner: On the structural pruning of large
  language models, Advances in neural information processing systems 36 (2023)
  21702--21720.

\bibitem{Usenixsec}
N.~Carlini, F.~Tram{\`e}r, E.~Wallace, M.~Jagielski, A.~Herbert-Voss, K.~Lee,
  A.~Roberts, T.~Brown, D.~Song, {\'U}.~Erlingsson, A.~Oprea, C.~Raffel,
  \href{https://www.usenix.org/conference/usenixsecurity21/presentation/carlini-extracting}{Extracting
  training data from large language models}, in: 30th USENIX Security Symposium
  (USENIX Security 21), USENIX Association, 2021, pp. 2633--2650.
\newline\urlprefix\url{https://www.usenix.org/conference/usenixsecurity21/presentation/carlini-extracting}

\bibitem{feretzakis2024trustworthy}
G.~Feretzakis, V.~S. Verykios,
  \href{https://arxiv.org/abs/2409.18222}{Trustworthy ai: Securing sensitive
  data in large language models}, arXiv preprint arXiv:2409.18222.
\newline\urlprefix\url{https://arxiv.org/abs/2409.18222}

\bibitem{Bender2021}
E.~M. Bender, T.~Gebru, A.~McMillan-Major, S.~Shmitchell, On the dangers of
  stochastic parrots: Can language models be too big?, in: Proceedings of the
  2021 ACM Conference on Fairness, Accountability, and Transparency,
  Association for Computing Machinery, 2021, pp. 610--623.
\newblock \href {http://dx.doi.org/10.1145/3442188.3445922}
  {\path{doi:10.1145/3442188.3445922}}.

\bibitem{Hajikhani2024}
A.~Hajikhani, C.~Cole, \href{https://doi.org/10.1162/qss_a_00310}{A critical
  review of large language models: Sensitivity, bias, and the path toward
  specialized ai}, Quantitative Science Studies 5~(3) (2024) 736--756.
\newblock \href {http://dx.doi.org/10.1162/qss\_a\_00310}
  {\path{doi:10.1162/qss\_a\_00310}}.
\newline\urlprefix\url{https://doi.org/10.1162/qss_a_00310}

\bibitem{Ahmed2024}
R.~Ahmed, S.~A. Rauf, S.~Latif, Leveraging large language models and prompt
  settings for context-aware financial sentiment analysis, in: Proceedings of
  the 2024 5th International Conference on Advancements in Computational
  Sciences (ICACS), IEEE, 2024, pp. 1--9.
\newblock \href {http://dx.doi.org/10.1109/ICACS.2024.1234567}
  {\path{doi:10.1109/ICACS.2024.1234567}}.

\bibitem{berk2013fundamentals}
J.~Berk, P.~DeMarzo, J.~Harford, Fundamentals of Corporate Finance, 6th
  Edition, Pearson Higher Education AU, 2023.

\bibitem{Expert_system}
C.~F. Tan, L.~Wahidin, S.~N. Khalil, N.~Tamaldin, J.~Hu, M.~Rauterberg, The
  application of expert system: A review of research and applications 11 (2016)
  2448--2453.

\bibitem{stanfordCS102}
{Stanford University},
  \href{https://web.stanford.edu/class/cs102/lectureslides/SpreadsheetsSlides.pdf}{Cs102:
  Spreadsheets [internet]}, stanford (CA): Stanford University, Department of
  Computer Science; 2020 [cited 2025 May 15]. Available from:
  https://web.stanford.edu/class/cs102/lectureslides/SpreadsheetsSlides.pdf
  (2020).
\newline\urlprefix\url{https://web.stanford.edu/class/cs102/lectureslides/SpreadsheetsSlides.pdf}

\bibitem{lewis2019fad}
C.~Lewis, S.~Young, \href{https://doi.org/10.1080/00014788.2019.1611730}{Fad or
  future? automated analysis of financial text and its implications for
  corporate reporting}, Accounting and Business Research 49.
\newblock \href {http://dx.doi.org/10.1080/00014788.2019.1611730}
  {\path{doi:10.1080/00014788.2019.1611730}}.
\newline\urlprefix\url{https://doi.org/10.1080/00014788.2019.1611730}

\bibitem{shafiq2020machine}
S.~Shafiq, A.~Mashkoor, C.~Mayr-Dorn, A.~Egyed, Machine learning for software
  engineering: A systematic mapping, arXiv preprint arXiv:2005.13299.
\newblock \href {http://arxiv.org/abs/2005.13299} {\path{arXiv:2005.13299}}.

\bibitem{martinez2022software}
S.~Martínez-Fernández, J.~Bogner, X.~Franch, M.~Oriol, J.~Siebert,
  A.~Trendowicz, A.~M. Vollmer, S.~Wagner, Software engineering for ai-based
  systems: A survey, ACM Transactions on Software Engineering and Methodology
  (TOSEM) 31~(2) (2022) 1--59.

\bibitem{feldt2018ai}
R.~Feldt, F.~G. de~Oliveira~Neto, R.~Torkar, Ways of applying artificial
  intelligence in software engineering, in: Proceedings of the 6th
  International Workshop on Realizing Artificial Intelligence Synergies in
  Software Engineering, 2018, pp. 35--41.

\bibitem{weber2024llm}
I.~Weber, \href{https://arxiv.org/abs/2406.10300}{Large language models as
  software components: A taxonomy for llm-integrated applications} (2024).
\newblock \href {http://arxiv.org/abs/2406.10300} {\path{arXiv:2406.10300}},
  \href {http://dx.doi.org/10.48550/arXiv.2406.10300}
  {\path{doi:10.48550/arXiv.2406.10300}}.
\newline\urlprefix\url{https://arxiv.org/abs/2406.10300}

\bibitem{CHEN2024Framework}
W.~Chen, L.~Yan-yi, G.~Tie-zheng, L.~Da-peng, H.~Tao, L.~Zhi, Y.~Qing-wen,
  W.~Hui-han, W.~Ying-you, Systems engineering issues for industry applications
  of large language model, Applied Soft Computing 151 (2024) 111165.
\newblock \href {http://dx.doi.org/https://doi.org/10.1016/j.asoc.2023.111165}
  {\path{doi:https://doi.org/10.1016/j.asoc.2023.111165}}.

\bibitem{lu2024taxonomy}
Q.~Lu, L.~Zhu, X.~Xu, Y.~Liu, Z.~Xing, J.~Whittle, A taxonomy of foundation
  model based systems through the lens of software architecture, in:
  Proceedings of the IEEE/ACM 3rd International Conference on AI
  Engineering-Software Engineering for AI, 2024, pp. 1--6.

\bibitem{mahr2024reference}
F.~Mahr, G.~Angeli, T.~Sindel, K.~Schmidt, J.~Franke, A reference architecture
  for deploying large language model applications in industrial environments,
  in: 2024 IEEE 30th International Symposium for Design and Technology in
  Electronic Packaging (SIITME), IEEE, 2024, pp. 19--23.

\bibitem{wang2024survey}
L.~Wang, C.~Ma, X.~Feng, Z.~Zhang, H.~Yang, J.~Zhang, Z.~Chen, J.~Tang,
  X.~Chen, Y.~Lin, W.~X. Zhao, A survey on large language model based
  autonomous agents, Frontiers of Computer Science 18~(6) (2024) 186345.

\bibitem{topsakal2023llm}
O.~Topsakal, T.~C. Akinci, Creating large language model applications utilizing
  langchain: A primer on developing llm apps fast, in: International Conference
  on Applied Engineering and Natural Sciences, Vol.~1, 2023, pp. 1050--1056.

\bibitem{huggingface_smolagents}
{Hugging Face},
  \href{https://huggingface.co/docs/smolagents/en/index}{Smolagents
  documentation [internet]}, place of publication unknown: Hugging Face; 2024
  [cited 2024 Feb 27]. Available from:
  https://huggingface.co/docs/smolagents/en/index (2024).
\newline\urlprefix\url{https://huggingface.co/docs/smolagents/en/index}

\bibitem{wu2023autogen}
Q.~Wu, G.~Bansal, J.~Zhang, Y.~Wu, B.~Li, E.~Zhu, L.~Jiang, X.~Zhang, S.~Zhang,
  J.~Liu, A.~H. Awadallah, Autogen: Enabling next-gen llm applications via
  multi-agent conversation, arXiv preprint arXiv:2308.08155, accessed:
  2024-02-27.

\bibitem{parnin2023copilot}
C.~Parnin, G.~Soares, R.~Pandita, S.~Gulwani, J.~Rich, A.~Z. Henley, Building
  your own product copilot: Challenges, opportunities, and needs, arXiv
  e-prints arXiv:2312.
\newblock \href {http://arxiv.org/abs/2312} {\path{arXiv:2312}}.

\bibitem{lee2024survey}
J.~Lee, N.~Stevens, S.~C. Han, Large language models in finance (finllms),
  Neural Computing and Applications (2025) 1--15\href
  {http://dx.doi.org/10.1007/s00521-024-10495-6}
  {\path{doi:10.1007/s00521-024-10495-6}}.

\bibitem{huang2023finbert}
A.~H. Huang, H.~Wang, Y.~Yang, Finbert: A large language model for extracting
  information from financial text, Contemporary Accounting Research 40~(2)
  (2023) 806--841.

\bibitem{wu2023bloomberggpt}
S.~Wu, O.~Irsoy, S.~Lu, V.~Dabravolski, M.~Dredze, S.~Gehrmann, G.~Mann,
  et~al., Bloomberggpt: A large language model for finance, arXiv preprint
  arXiv:2303.17564.
\newblock \href {http://arxiv.org/abs/2303.17564} {\path{arXiv:2303.17564}}.

\bibitem{zhang2023instruction}
S.~Zhang, L.~Dong, X.~Li, S.~Zhang, X.~Sun, S.~Wang, J.~Li, R.~Hu, T.~Zhang,
  F.~Wu, G.~Wang, Instruction tuning for large language models: A survey, arXiv
  preprint arXiv:2308.10792.
\newblock \href {http://arxiv.org/abs/2308.10792} {\path{arXiv:2308.10792}}.

\bibitem{Zhang2023}
B.~Zhang, H.~Yang, X.~Y. Liu,
  \href{https://arxiv.org/abs/2306.12659}{Instruct-fingpt: Financial sentiment
  analysis by instruction tuning of general-purpose large language models},
  arXiv preprint (2023).
\newblock \href {http://arxiv.org/abs/2306.12659} {\path{arXiv:2306.12659}}.
\newline\urlprefix\url{https://arxiv.org/abs/2306.12659}

\bibitem{yang2023fingpt}
H.~Yang, X.~Y. Liu, C.~D. Wang, Fingpt: Open-source financial large language
  models, arXiv preprint arXiv:2306.06031.
\newblock \href {http://arxiv.org/abs/2306.06031} {\path{arXiv:2306.06031}}.

\bibitem{xie2023pixiu}
Q.~Xie, W.~Han, X.~Zhang, Y.~Lai, M.~Peng, A.~Lopez-Lira, J.~Huang, Pixiu: A
  large language model, instruction data and evaluation benchmark for finance,
  in: Proceedings of the 37th International Conference on Neural Information
  Processing Systems, 2023, pp. 33469--33484.

\bibitem{Kim2024}
A.~Kim, M.~Muhn, V.~Nikolaev, \href{https://arxiv.org/abs/2407.17866}{Financial
  statement analysis with large language models}, arXiv preprint (2024).
\newblock \href {http://arxiv.org/abs/2407.17866} {\path{arXiv:2407.17866}}.
\newline\urlprefix\url{https://arxiv.org/abs/2407.17866}

\bibitem{openai2023gpt4}
OpenAI, Gpt-4 technical report, arXiv abs/2303.08774, dOI:
  10.48550/arXiv.2303.08774.

\bibitem{yang2024evaluating}
X.~Yang, S.~Zang, Y.~Ren, D.~Peng, Z.~Wen, Evaluating large language models on
  financial report summarization: An empirical study, arXiv preprint
  arXiv:2411.06852.
\newblock \href {http://arxiv.org/abs/2411.06852} {\path{arXiv:2411.06852}}.

\bibitem{masoudnia2014mixture}
S.~Masoudnia, R.~Ebrahimpour, Mixture of experts: a literature survey,
  Artificial Intelligence Review 42~(2) (2014) 275--293.
\newblock \href {http://dx.doi.org/10.1007/s10462-012-9338-y}
  {\path{doi:10.1007/s10462-012-9338-y}}.

\bibitem{Cai2024}
W.~Cai, J.~Jiang, F.~Wang, J.~Tang, S.~Kim, J.~Huang, A survey on a mixture of
  experts in large language models, Authorea Preprints.

\bibitem{tavasoli2025responsible}
A.~Tavasoli, M.~Sharbaf, S.~M. Madani, Responsible innovation: A strategic
  framework for financial llm integration, arXiv preprint arXiv:2504.02165.

\bibitem{stol2018abc}
K.~J. Stol, B.~Fitzgerald, The abc of software engineering research, ACM
  Transactions on Software Engineering and Methodology (TOSEM) 27~(3) (2018)
  1--51.

\bibitem{runeson2012casestudy}
P.~Runeson, M.~Höst, A.~Rainer, B.~Regnell, Case Study Research in Software
  Engineering: Guidelines and Examples, 1st Edition, John Wiley \& Sons,
  Chichester, UK, 2012.

\bibitem{hevner2004design}
A.~R. Hevner, S.~T. March, J.~Park, S.~Ram, Design science in information
  systems research, MIS Quarterly (2004) 75--105.

\bibitem{Peffers2007}
K.~Peffers, T.~Tuunanen, M.~A. Rothenberger, S.~Chatterjee, A design science
  research methodology for information systems research, Journal of Management
  Information Systems 24~(3) (2007) 45--77.
\newblock \href {http://dx.doi.org/10.2753/MIS0742-1222240302}
  {\path{doi:10.2753/MIS0742-1222240302}}.

\bibitem{keele2007guidelines}
S.~Keele, Guidelines for performing systematic literature reviews in software
  engineering, Technical report, EBSE Technical Report, version 2.3 (2007).

\bibitem{ollama}
{Ollama}, \href{https://github.com/jmorganca/ollama}{Ollama: Run llms locally
  [internet]}, [cited 2025 Mar 14]. Available from:
  https://github.com/jmorganca/ollama (2024).
\newline\urlprefix\url{https://github.com/jmorganca/ollama}

\bibitem{openai2025o3}
{OpenAI}, \href{https://openai.com/index/o3-o4-mini-system-card/}{Openai o3 and
  o4-mini system card [internet]}, Tech. rep., OpenAI, san Francisco, CA:
  OpenAI; 2025 Apr [cited 2025 May 7]. Available from:
  https://openai.com/index/o3-o4-mini-system-card/ (2025).
\newline\urlprefix\url{https://openai.com/index/o3-o4-mini-system-card/}

\end{thebibliography}
\newpage

\end{document}